\documentclass[sn-mathphys,Numbered]{sn-jnl}

\usepackage{graphicx}%
\usepackage{multirow}%
\usepackage{amsmath,amssymb,amsfonts}%
\usepackage{amsthm}%
\usepackage{mathrsfs}%
\usepackage[title]{appendix}%
\usepackage{xcolor}%
\usepackage{textcomp}%
\usepackage{manyfoot}%
\usepackage{booktabs}%
\usepackage{algorithm}%
\usepackage{algorithmicx}%
\usepackage{algpseudocode}%
\usepackage{listings}%
\usepackage{accents}
\usepackage{pdfpages}
\usepackage{booktabs}  
\usepackage{caption}   
\usepackage{array}     
\usepackage[absolute,overlay]{textpos}
\usepackage{bm}
\usepackage{amsmath}
\usepackage{graphicx}
\usepackage{subcaption}
\usepackage{booktabs}
\usepackage{multirow}
\usepackage{makecell}
\usepackage{array}
\newcolumntype{P}[1]{>{\raggedright\arraybackslash}p{#1}}
\usepackage{enumitem}
\usepackage{booktabs} 
\usepackage{caption} 
\usepackage{float} 
\usepackage{mwe}
\theoremstyle{thmstyleone}%

\theoremstyle{thmstyletwo}%

\theoremstyle{thmstylethree}%

\begin{document}

\title[Article Title]{Introducing a microstructure-embedded autoencoder approach for reconstructing high-resolution solution field data from a reduced parametric space}


\author*[1]{\fnm{Rasoul} \sur{Najafi Koopas}}\email{najafikr@hsu-hh.de}

\author[2]{\fnm{Shahed} \sur{Rezaei}}\email{s.rezaei@access-technology.de}

\author[1]{\fnm{Natalie} \sur{Rauter}}\email{natalie.rauter@hsu-hh.de}

\author[1]{\fnm{Richard} \sur{Ostwald}}\email{ostwald@hsu-hh.de}
\author[1]{\fnm{Rolf} \sur{Lammering}}\email{rolf.lammering@hsu-hh.de}

\affil*[1]{\orgdiv{Institute of Solid Mechanics}, \orgname{Helmut-Schmidt University/University of the Federal Armed Forces}, \orgaddress{\street{Holstenhofweg  85}, \city{Hamburg}, \postcode{22043}, \country{Germany}}}

\affil[2]{Access e.V., Aachen, Germany}

\abstract{
In this study, we develop a novel multi-fidelity deep learning approach that transforms low-fidelity solution maps into high-fidelity ones by incorporating parametric space information into a standard autoencoder architecture. This method's integration of parametric space information significantly reduces the need for training data to effectively predict high-fidelity solutions from low-fidelity ones. In this study, we examine a two-dimensional steady-state heat transfer analysis within a highly heterogeneous materials microstructure. The heat conductivity coefficients for two different materials are condensed from a $101 \times 101$ grid to smaller grids. We then solve the boundary value problem on the coarsest grid using a pre-trained physics-informed neural operator network known as Finite Operator Learning (FOL). The resulting low-fidelity solution is subsequently upscaled back to a $101 \times 101$ grid using a newly designed enhanced autoencoder. The novelty of the developed enhanced autoencoder lies in the concatenation of heat conductivity maps of different resolutions to the decoder segment in distinct steps. Hence the developed algorithm is named microstructure-embedded autoencoder (MEA). We compare the MEA outcomes with those from finite element methods, the standard U-Net, and various other upscaling techniques, including interpolation functions and feedforward neural networks (FFNN). Our analysis shows that MEA outperforms these methods in terms of computational efficiency and error on test cases. As a result, the MEA serves as a potential supplement to neural operator networks, effectively upscaling low-fidelity solutions to high fidelity while preserving critical details often lost in traditional upscaling methods, particularly at sharp interfaces like those seen with interpolation. 
}




\maketitle
\newpage
\section{Introduction}\label{sec1}

High-fidelity simulations of physical phenomena are critically dependent on the balance between computational complexity and the accuracy of high-resolution models. In computational mechanics, where precision is essential, the trade-off between computational efficiency and the level of detail in the solution fields poses a significant challenge. High-fidelity models, while offering detailed insights, are computationally intensive and demand substantial resources in terms of both time and computational power. Conversely, low-resolution models, which are less demanding computationally, often lack important details, compromising the accuracy necessary in fields where precision is essential. This challenge highlights the necessity of optimizing both the performance and the cost of computational simulations in scientific research.

To address this challenge in both, scientific research and application, the multi-fidelity modeling method \cite{peherstorfer2018survey, fernandez2016review} is considered a potential approach to effectively reduce the acquisition cost of high-fidelity data. Accurate but costly, high-fidelity data is typically obtained from detailed experiments or simulations. Conversely, more affordable low-fidelity data, though less accurate, offers a basic representation of physical phenomena. The multi-fidelity approach aims to combine low and high-fidelity data to improve both the cost-effectiveness and accuracy of the physical phenomena under consideration. Traditional multi-fidelity modeling approaches, such as Co-Kriging \cite{le2014recursive, perdikaris2015multi} and multi-level Monte Carlo \cite{giles2008multilevel, bierig2016approximation}, have proven effective in reducing the acquisition costs of high-fidelity data. 

The use of deep neural networks (DNNs) for multi-fidelity modeling takes advantage of their exceptional ability to generalize. The integration of deep neural networks to construct multi-fidelity architectures has become a prominent research topic in recent years. Interested readers are referred to works such as \cite{minisci2011robust, he2020multi, zhang2021multi, chen2022multi, song2022transfer, meng2020composite}, focusing on the development of deep neural network architectures. These architectures aim to reduce computational costs for generating high-fidelity solutions by incorporating both low- and high-fidelity data. A common issue with the aforementioned methods is their dependence on multiple neural networks for multi-fidelity modeling, necessitating the integration of various networks for practical inference. This approach often results in redundant structures within the serial multi-fidelity model and can lead to error accumulation and propagation across the series of models. Furthermore, they frequently overlook the physical principles fundamental to engineering within their frameworks. To address this deficiency and reduce the number of labeled data for training, efforts like those cited in \cite{chakraborty2021transfer, aliakbari2022predicting, zhang2023multi, shu2023physics} employ Physics-Informed Neural Networks (PINN), which utilize physical insights to broaden the multi-fidelity modeling scope. Although promising, integrating PINNs into multi-fidelity models is challenging. Up until now, these models require complex tuning due to the increased number of hyperparameters within their loss function, making their training costly and time-consuming. Therefore, utilizing PINNs for solving partial differential equations (PDEs) in a parametric way and for a class of problems with realistic engineering applications is hindered and shall be improved.

Another promising architecture for mapping two different fields is the so-called U-Net method. The U-Net architecture, originally developed for medical image segmentation \cite{ronneberger2015u}, has attracted great interest in numerous disciplines, such as computational mechanics \cite{gupta2023accelerated, sepasdar2021data, yan2023multi}, due to its superior performance and adaptability. Known for its efficient management of complex spatial hierarchies, this deep learning framework has demonstrated remarkable versatility and extended its utility beyond its original healthcare context.

It is also worth mentioning that physics-informed operator learning techniques, which can be utilized in handling reduced parameter spaces, are increasingly recognized for their benefits (as investigated in \cite{Wang_Paris2021, li2023physicsinformed, Faroughi2023} and reviewed in \cite{rezaei2024integration}). These methods aim to train networks to understand underlying physical equations \textbf{parametrically}, applicable across various boundary value problems rather than a specific boundary value problem \cite{Wang_Paris2021, Liu2024}. This approach extends the principles of PINNs, initially introduced by \citet{RAISSI2019}. To improve efficiency and accuracy of physics-informed operator learning techniques, \citet{kontolati2023learning} suggested mapping high-dimensional datasets to a low-dimensional latent space using specialized autoencoders and pre-trained decoders. 

To address the challenges associated with multi-fidelity approaches, we introduced a novel architecture called the Microstructure-Embedded Autoencoder (MEA). In this architecture, the low-fidelity solution map is first obtained by a solver on the reduced parametric space. Afterwards, the resulting low-fidelity solution is then upscaled to a high-fidelity map via an enhanced autoencoder. 

The innovation of the MEA model is twofold. First, unlike many architectures that rely on picture data for computations, MEA uses the spatial distribution of material properties as input, thereby indirectly incorporating physical influences into the deep neural network architecture. Second, instead of employing multiple networks and inference processes, MEA uses a single autoencoder to enhance the resolution of the solution map. 

In this work, we have used a finite operator learning (FOL) architecture to solve the boundary value problem, which is steady-state heat transfer analysis, in reduced parameter space. This approach eliminates problems commonly associated with other solvers, such as those specific to a boundary value problem, and increases the versatility of MEA model.

\begin{figure}[t]
\center
    \includegraphics[width=\textwidth]{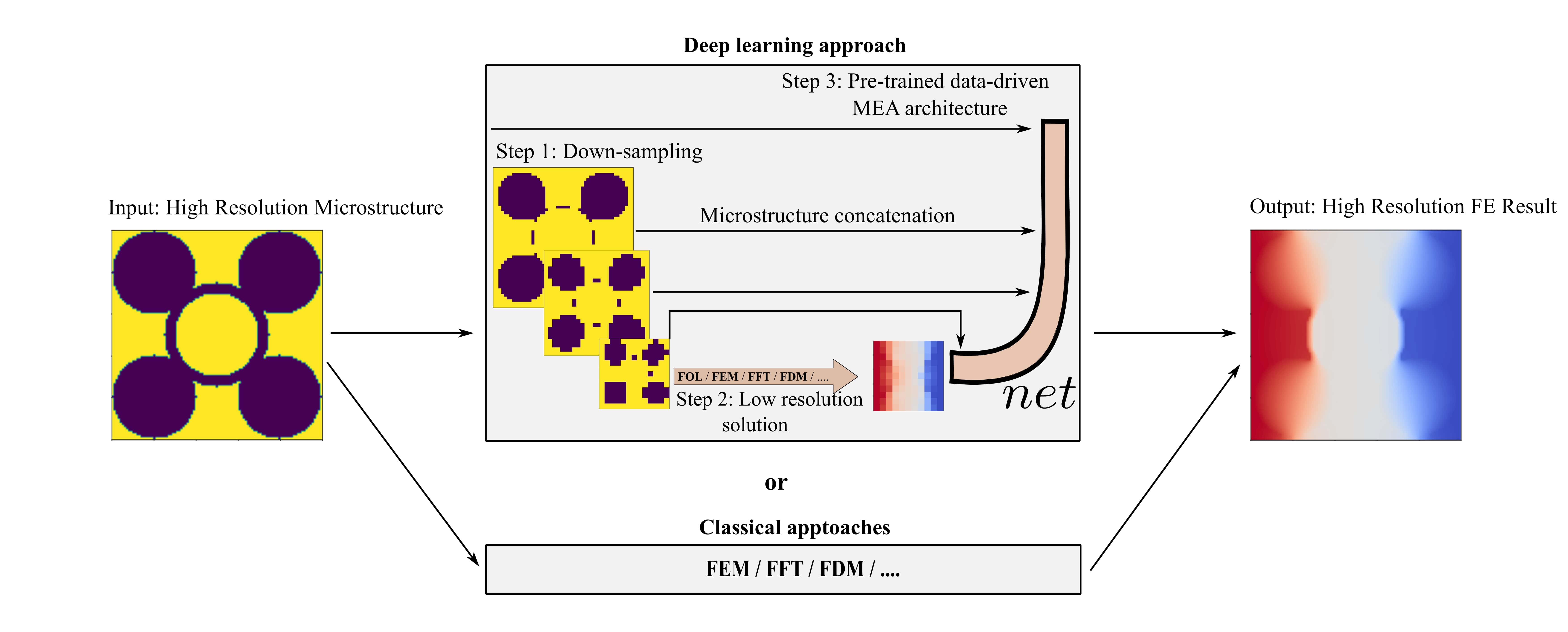}
    \caption{In the developed MEA architecture, the initial step involves condensing a high-resolution heat conductivity map into various lower-resolution maps. In the next step, the coarsest grid—defined in this study as an \(11 \times 11\) grid—is utilized to solve the boundary value problem. For this task, one of several techniques, including the Finite Element Method (FEM), Finite Difference Method (FDM), Fast Fourier Transform (FFT), Finite Operator Learning (FOL), or Physics-Informed Neural Networks (PINNs), among others can be implemented. Following the resolution of the boundary value problem at this lower scale, the resultant low-resolution output undergoes an upscaling process through the use of an enhanced autoencoder. }
    \label{Introduction}
\end{figure}
As can be seen in Fig.~\ref{Introduction}, within the MEA architecture, detailed high-resolution fields are reconstructed by integrating the spatial distribution of material properties of different resolutions in different stages of the decoder segment. The general MEA architecture involves a three-step process: first, it starts with a condensation process to generate material property distribution maps of different grid sizes, second, solving the boundary value problem on the coarsest grid with an available numerical solver (FEM, FDM, PINNs, FOL, ...), and finally, scaling to a high-resolution solution map using an enhanced autoencoder architecture. As shown in this work, this method significantly reduces the calculation time compared to traditional methods such as the finite element method, making it both efficient and accurate for computational mechanics. 
Moreover, to demonstrate the power and robustness of the MEA approach, we conduct a comparative analysis against several established methods. These include the standard U-Net architecture, as well as other techniques that utilize interpolation functions to upscale the solution. 
The comparative results indicate the superior performance of MEA in terms of accuracy and computational cost.


The structure of this paper is organized as follows: Section 2 describes the boundary value problem employed in this study for data generation and outlines the methodologies used to solve the heat transfer problem, specifically through the finite element method and finite operator learning as numerical solvers. Section 3 details the architecture of the algorithms used to obtain high-fidelity solution maps. These algorithms include the standard U-Net architecture, the microstructure-embedded autoencoder, interpolation techniques, and feed-forward neural networks. In Section 4, following a comprehensive hyperparameter study, the performance of the MEA approach is evaluated against other established methods in the literature. Finally, the paper concludes by summarizing the main results.


\newpage
\section{Problem description} \
\subsection{Necessity of reducing parameter space}
\begin{figure}[H] 
  \centering
    \includegraphics[width=0.99\linewidth]{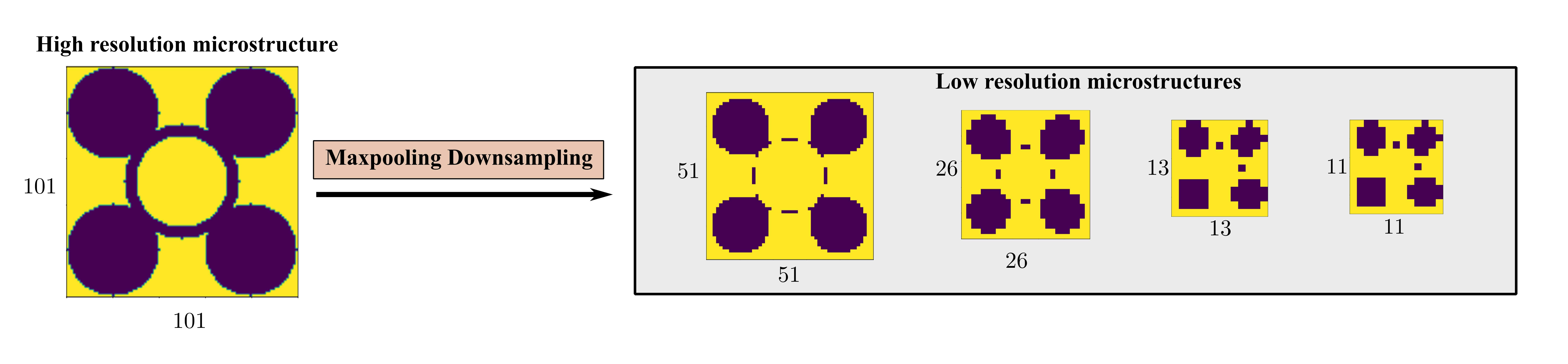}
\caption{Condensation of high-resolution parametric space into lower-resolution parametric spaces using the \texttt{MaxPool} function from the \texttt{SciPy} library.}
  \label{maxpool}
\end{figure}

Regardless of the numerical method used to solve a given boundary value problem parametrically, it is always beneficial from a computational cost standpoint to reduce the parameter space as much as possible. This reduction facilitates a faster training phase for operator learning algorithms. Moreover, working with a fewer number of inputs is not only more convenient but also easier to interpret. Therefore, we propose a simple approach to reduce the parametric space (microstructure resolution) for our problem. This approach is inspired by algorithms used in convolutional neural networks and is illustrated in Fig.~\ref{maxpool}.
The reduction of the parametric space involves transforming the high-resolution material property map into lower-resolution representations through \texttt{Max Pooling} operations. \texttt{Max Pooling} is a process commonly used in image processing and deep learning for reducing the spatial size of an image or matrix while preserving the most significant features, such as edges or textures. This technique involves sliding a window across the input matrix and replacing the output's value with the maximum value within that window. The size of the window determines the degree of downsampling where larger windows result in more significant reduction and coarser representations of the original matrix. In this study, the high-resolution parametric space is condensed into four different resolutions, namely: \(N_{\text{51R}} = 51\), \(N_{\text{26R}} = 26\), \(N_{\text{13R}} = 13\), and \(N_{\text{11R}} = 11\).
\subsection{Problem statement}
\begin{figure}[t]
\center
    \includegraphics[width=1\textwidth]{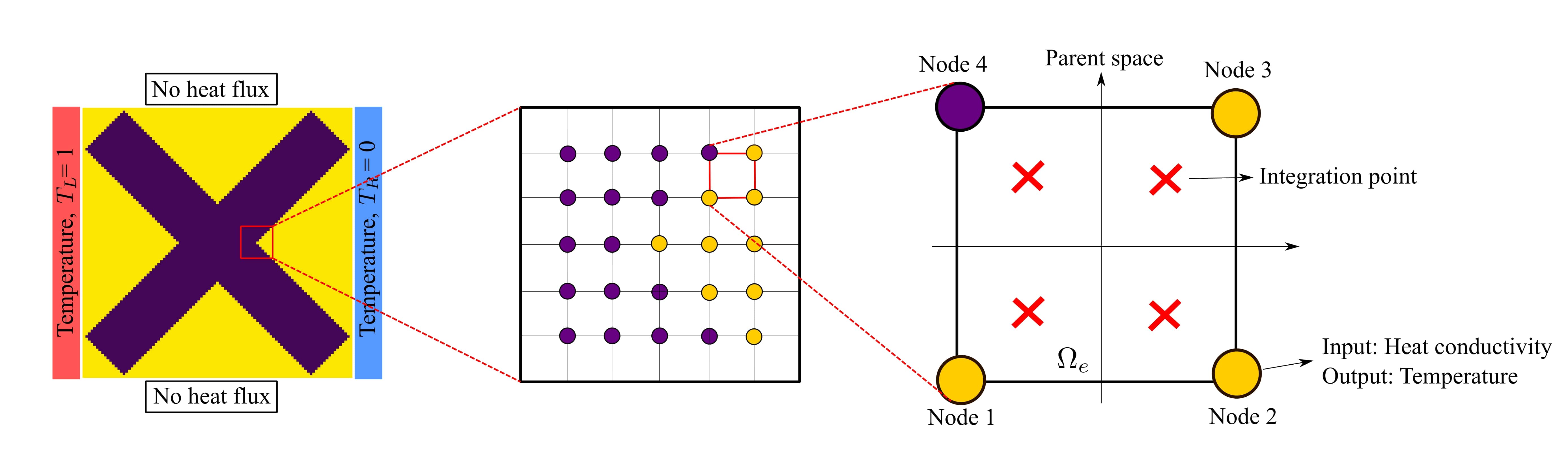}
    \caption{Description of problem setup, boundary conditions, and finite element meshes. The information of the domain is conveyed through a finite number of sensor points or nodes.}
    \label{BCs}
\end{figure}

As shown in Fig.~\ref{BCs}, this study addresses a two-dimensional steady-state heat transfer problem in a heterogeneous material. Same derivation principles are applied to analogous problems goverened by Poisson's equation, such as electrostatics, electric conduction, magnetostatic problems \cite{Guo2022}, chemical diffusion \cite{REZAEI2021104612}, and Darcy flow \cite{Goswami2023} under some assumptions.

In the current work, the parametric input space consists of the two-dimensional spatial distribution of heat conductivity coefficients. Moreover, the solution space represents the temperature field throughout the microstructure. The analysis domain, as depicted on the left side of Fig.~\ref{BCs}, is discretized into equally shaped quadrilateral finite element meshes. Additionally, boundary conditions, including Neumann (flux-free) and Dirichlet (with fixed temperature), are illustrated in Fig.~\ref{BCs}. The two material phases, differentiated by distinct colors, are homogeneous within their respective phase.


The heat flux vector $\bm{q}$ is modelled according to Fourier's assumption, 
\begin{align}
\label{Fourier}
\bm{q} = -k(x,y)\,\nabla T. 
\end{align} 
Here $k(x,y)$ is the phase-dependent heat conductivity coefficient and $\nabla T$ represents the spatial gradient of the temperature field $T$. The governing equation for this problem reads
\begin{align}
\label{StrongfromThermal}
\text{div}(\bm{q}) + Q &= 0~~~~~~ \text{in}~ \Omega, \\
\label{BCsthermalD}
T &= \bar{T}~~~~~\text{on}~ \Gamma_D, \\
\label{BCsthermalN}
\bm{q}\cdot \bm{n} = q_n &= \bar{q}~~~~~~\text{on}~ \Gamma_N.
\end{align} 
In the above relations, $Q$ is the heat source term. Moreover, the Dirichlet and Neumann boundary conditions are introduced in Eq.~\ref{BCsthermalD} and Eq.~\ref{BCsthermalN}, respectively. 

\subsection{Finite Element Modeling}
By introducing $\delta T$ as a test function and with integration by parts, the weak form of the steady-state diffusion problem reads
\begin{align}
\label{eq:weakformthermal}
\int_{\Omega}\,k(x,y)\,\nabla^T T\,\delta(\nabla T)~\mathrm{d}V\,+\,\int_{\Gamma_N}\bar{q}~\delta T~\mathrm{d}A\,-\int_{\Omega}\,Q\,\delta T~\mathrm{d}V\,=\,0.
\end{align}
Following the standard finite element method, the temperature field $T$, conductivity field $k$ as well as their first spatial derivatives, are approximated as
\begin{equation}
T= \sum {N_T}_i T_i =\boldsymbol{N}_{T} \boldsymbol T_e, \quad \nabla T =\sum {B_T}_i T_i =\boldsymbol{B}_{T}\boldsymbol T_e, \quad 
k= \sum {N_T}_i k_i =\boldsymbol{N}_{T} \boldsymbol k_e. 
\end{equation}
Here, $T_i$ and $k_i$ are the nodal values of the temperature and conductivity field of node $i$ of element $e$, respectively. Matrices $\boldsymbol N_T$ and $\boldsymbol B_T$ store the corresponding shape functions and their spatial derivatives \cite{REZAEI2022PINN, rezaei2024integration}. To compute these derivatives, we utilize the Jacobian matrix $\boldsymbol J = \partial \boldsymbol X / \partial \boldsymbol \xi$, where $\boldsymbol X = [x, y]$ and $\boldsymbol \xi = [\xi, \eta]$ represent the physical and parent coordinate systems, respectively \cite{bathe, hughes}.
Finally, one can write the discretized version of the weak form for one general finite element as\begin{align}
\label{eq:dis_residual}
\boldsymbol r_{eT} = \int_{\Omega_e} [\boldsymbol B_{T}]^T k~[\boldsymbol B_{T} \boldsymbol T_e] ~\mathrm{d}V - \int_{\Omega_t}[\boldsymbol  N_{T}]^T k~[\boldsymbol B \boldsymbol T_e]^T~\boldsymbol  n~\mathrm{d}S -\int_{\Omega} [\boldsymbol  N_{T}]^T  Q~\mathrm{d}V.
\end{align}

\subsection{Finite operator learning}
Motivated by the original formulation of the FEM and the concept of PINNs \cite{RAISSI2019}, in a recent study \cite{rezaei2024integration}, the authors proposed to construct the neural network's loss function based on the governing physical equations derived from the discretized weak form of the problem. Consequently, one can train the deep learning model in a parametric manner and utilize ideas from operator learning to map the physical input field to the solution field by training the deep learning model on a feasible amount of collocation fields (denoted as $k(x,y)$ in this work). It is demonstrated that by adopting this approach, one can achieve accuracy comparable to that of the finite element method while being computationally faster, as it only requires evaluating the network even for unseen cases \cite{rezaei2024integration}.

Based on the above introduction to the method, the output layer $\bm{Y}={T_i}$ consists of the components of the temperature field at each node. One can employ separate or fully connected feed-forward neural networks for each output variable. The neural network model is summarized as
\begin{align}
\label{eq:in_out}  
    T_{i} = \mathcal{N}_{i} (\bm{X}; \bm{\theta}_i),~~~\bm{X}=\{k_j\},~~~\bm{\theta}_i=\{\bm{W}_i,\bm{b}_i\},~~~i,j = 1 \cdots N.
\end{align}
Here, the trainable parameters of the $i$-th network are denoted by $\bm{\theta}_i$. Moreover, $\bm{W}_i$ and $\bm{b}_i$ encompass all the weights and biases within the $i$-th neural network. Additionally, $k_j$ represents the conductivity value at node $j$. The number of nodes is denoted by $N$.
\begin{figure}[t] 
  \centering
    \includegraphics[width=0.99\linewidth]{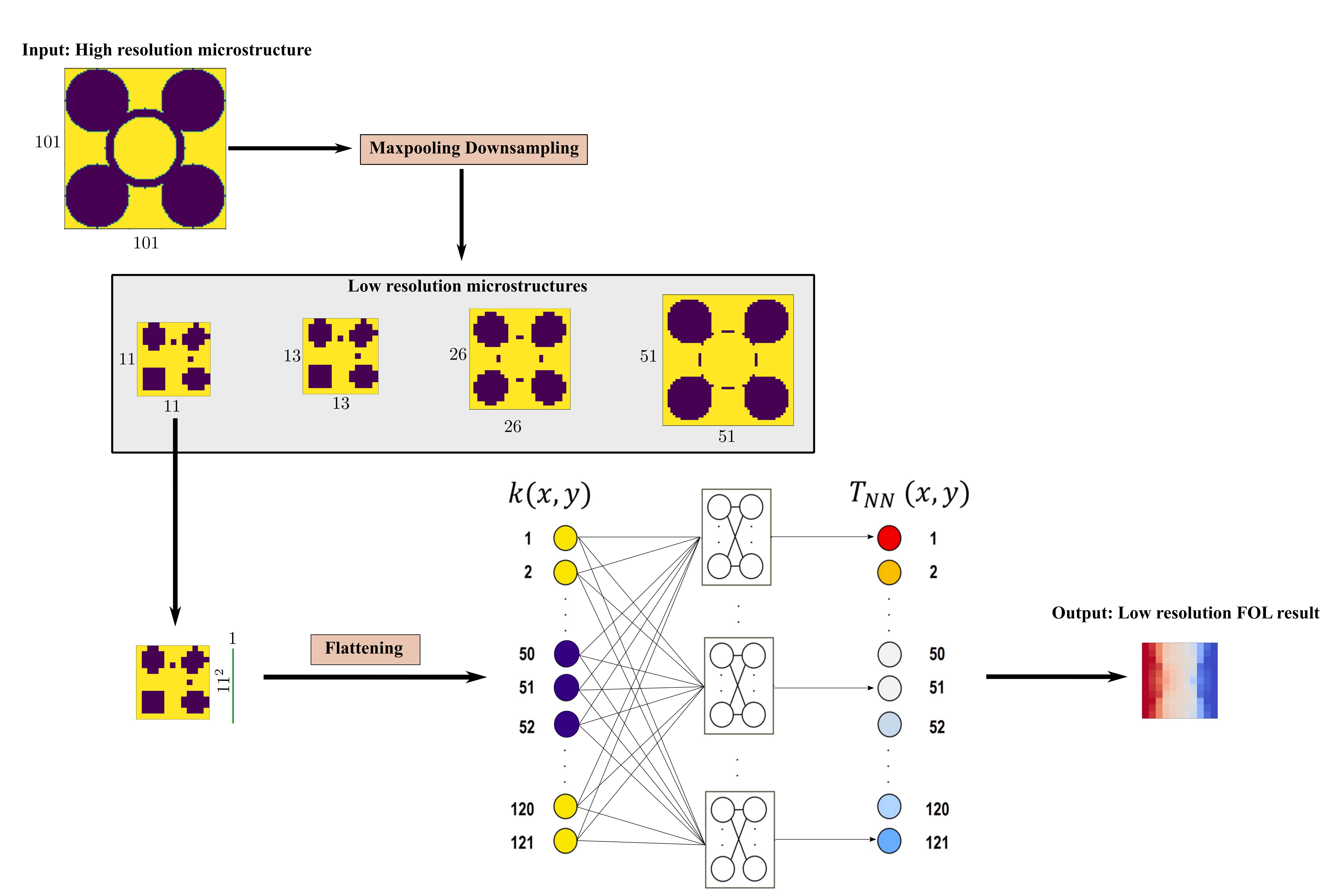}
  \caption{ Network architecture for finite operator learning, where information about the input parameter at each grid point goes in and the The output layer is evaluated by training the network based on the discretized weak form. By satisfying the residual or energy form of the problem through the optimization process in the deep learning model, physical outcomes or solutions are obtained for a parametric input space.}
  \label{fig:NN_idea}
\end{figure}
Next, we introduce the loss function where we assume no heat source term, i.e., $Q=0$. Second, we consider the upper and right edges as isolated, resulting in vanishing normal heat flux at the Neumann boundaries (see Fig.~\ref{BCs}). The temperature values at the left and right edges are fixed at $1.0$ and $0.0$ which are applied in a hard way in the network by omitting these nodes from the output layer. Finally, the total loss term $\mathcal{L}_{\text{tot}}$ which is implemented in Python using the JAX \cite{jax2018github} and Sciann \cite{SciANN} reads 
\begin{align}
\label{eq:loss_sum}
\mathcal{L}_{tot} =  
\lambda_e
\sum_{e=1}^{n_{el}}\boldsymbol T^T_e \left[ \boldsymbol \sum_{n=1}^{n_{int}} ~(\boldsymbol N_T \boldsymbol k_e)\boldsymbol B_{T}^T \boldsymbol B_{T} \right] \boldsymbol T_e.
\end{align}

As depicted in the upper right section of Fig.~\ref{BCs}, we utilize the positions of the standard four Gaussian points within the parent space to compute the energy integral. Moreover, we use quadrilateral elements with four integration points (i.e. $w_n=1$), where the determinant of the Jacobian matrix remains constant. Therefore, we define $\lambda_e=\dfrac{w_n}{2}\text{det}(\boldsymbol J)$. Fig.~\ref{fig:NN_idea} illustrates the architecture of the FOL used in this study to solve the boundary value problem.

\section{Sample generation}

In the development of our deep learning models for predicting the heat conductivity of two-phase materials, generating a diverse and representative training dataset is crucial. This dataset comprises spatial distributions that simulate 2D heat conductivity problems. The process begins by defining a square domain discretized into an \(N \times N\) grid, with each grid point representing a spatial coordinate in our simulated material sample. For each sample, elliptical inclusions with varying lengths of inner and outer axes and orientations are introduced, representing the two distinct phases of the material. The conductivities within these phases are determined by predefined values (\(k_{\mathrm{in}}\) and \(k_{\mathrm{out}}\)), which represent the thermal conductivities of the inner and outer phases, respectively.

\begin{figure}[H]
\center
    \includegraphics[width=0.99\textwidth]{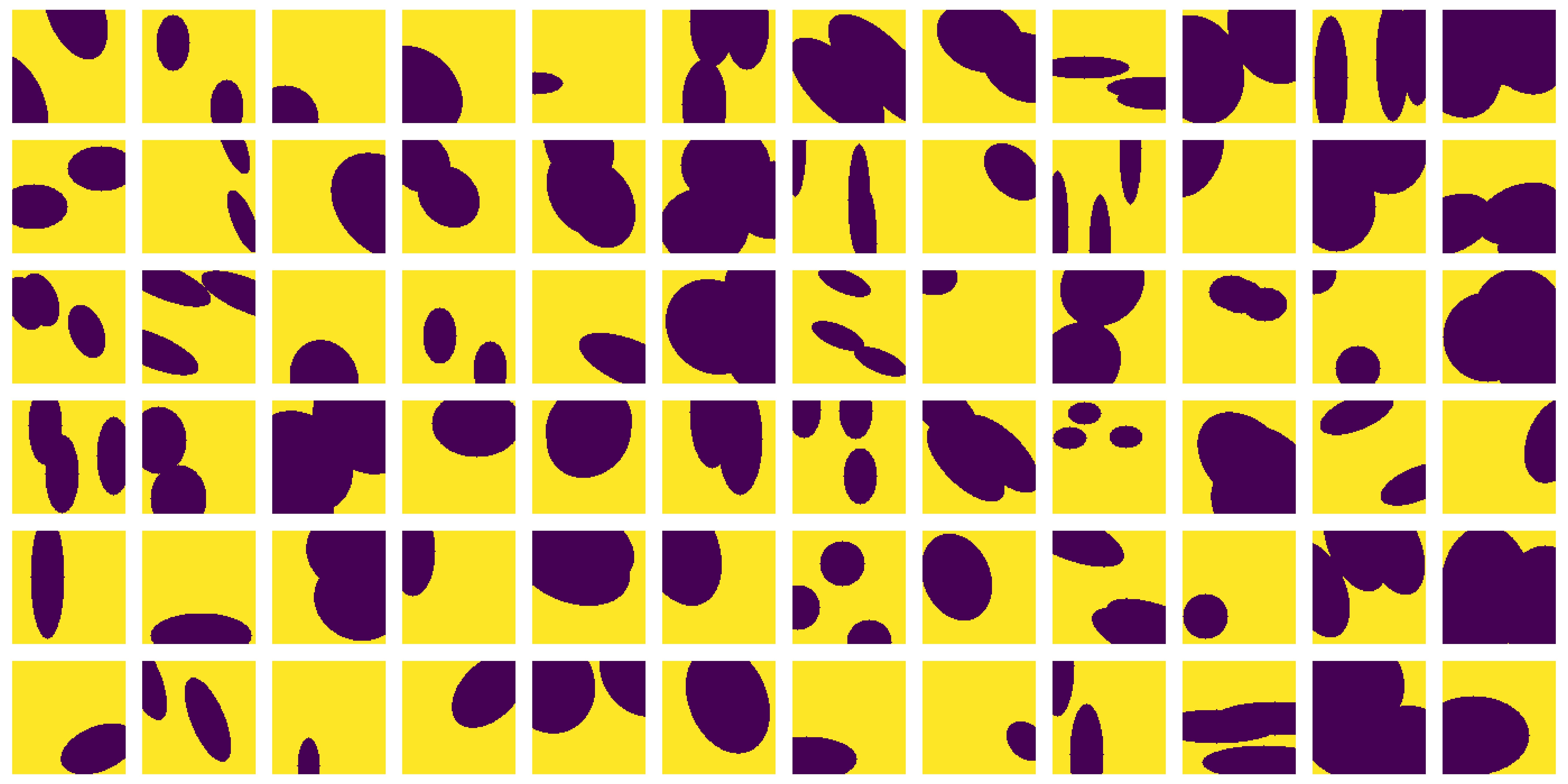}
    \caption{Randomly selected training samples from a dataset of 5,670 illustrate the spatial conductivity maps of two-phase materials within the training set. }
    \label{training_sample}
\end{figure}

To systematically investigate the space of possible microstructures, a parametric approach is employed where the number of ellipses (\(n_\mathrm{c}\)), the axes lengths of the ellipses (\(a_{\mathrm{outer}}, b_{\mathrm{outer}}, a_{\mathrm{inner}}, b_{\mathrm{inner}}\)), and the orientation (\(\theta\)) are varied according to predefined ranges and steps. This method allows for the generation of a comprehensive set of microstructural configurations. Each configuration's thermal conductivity distribution is computed using a spatial function, which calculates the conductivity at each point in the domain based on the defined microstructure. In total, about 5,670 samples are generated in high resolution (i.e. $N_{\mathrm{HR}}=101$), with some of them depicted in Fig.~\ref{training_sample} for the sake of illustration only. 

These datasets are then fed into a finite element solver to compute the corresponding temperature fields. The set of high-resolution input morphology (i.e., the spatial distribution of the heat conductivity coefficients) as well as high-resolution output temperatures are then used for the following three purposes:

\begin{itemize}
    \item The set of high-resolution input-output pairs is used for training the U-Net model, as will be described in Section \ref{U-Net Section}.
    \vspace{0.1cm}
    \item The high-resolution inputs, with \(N_{\text{HR}} = 101\), are condensed to lower resolutions of \(N_{\text{51R}} = 51\), \(N_{\text{26R}} = 26\), \(N_{\text{13R}} = 13\), and \(N_{\text{11R}} = 11\). Subsequently, the lowest resolution, \(N_{\text{11R}} = 11\), is used for unsupervised training of the finite operator learning model described in Section 2.4.
    \vspace{0.1cm}
    \item Finally, the high-resolution heat conductivity maps along with their corresponding condensed maps will be used for training the new microstructure-embedded autoencoder developed in this work.
\end{itemize}
\vspace{0.25cm}

\begin{figure}[H]
\center
    \includegraphics[width=0.99\textwidth]{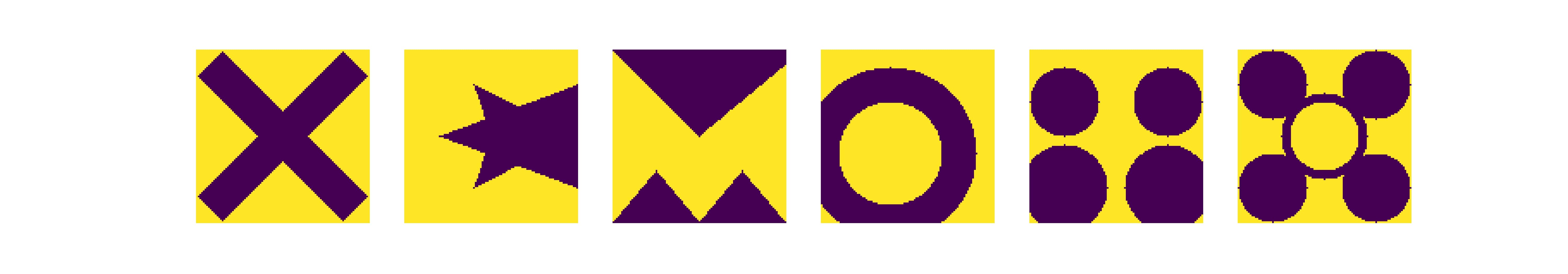}
    \caption{Out-of-Distribution test samples. These samples, not seen during training, test the model's ability to adapt to new, unseen cases.}
    \label{test_cases}
\end{figure}

Referring to Fig.~\ref{training_sample}, which showcases randomly selected samples from the training dataset, and considering Fig.~\ref{test_cases} showing the test samples, it is evident that some of the test cases significantly diverge from the training distribution, categorizing them as out-of-distribution (OOD). These OOD test cases represent scenarios substantially different from those observed during training (e.g. there are no ring, triangular, or rectangular shapes in the training set). This deliberate selection aims to rigorously assess the robustness and generalization capabilities of the neural networks.

\begin{figure}[H]
\center
    \includegraphics[width=\textwidth]{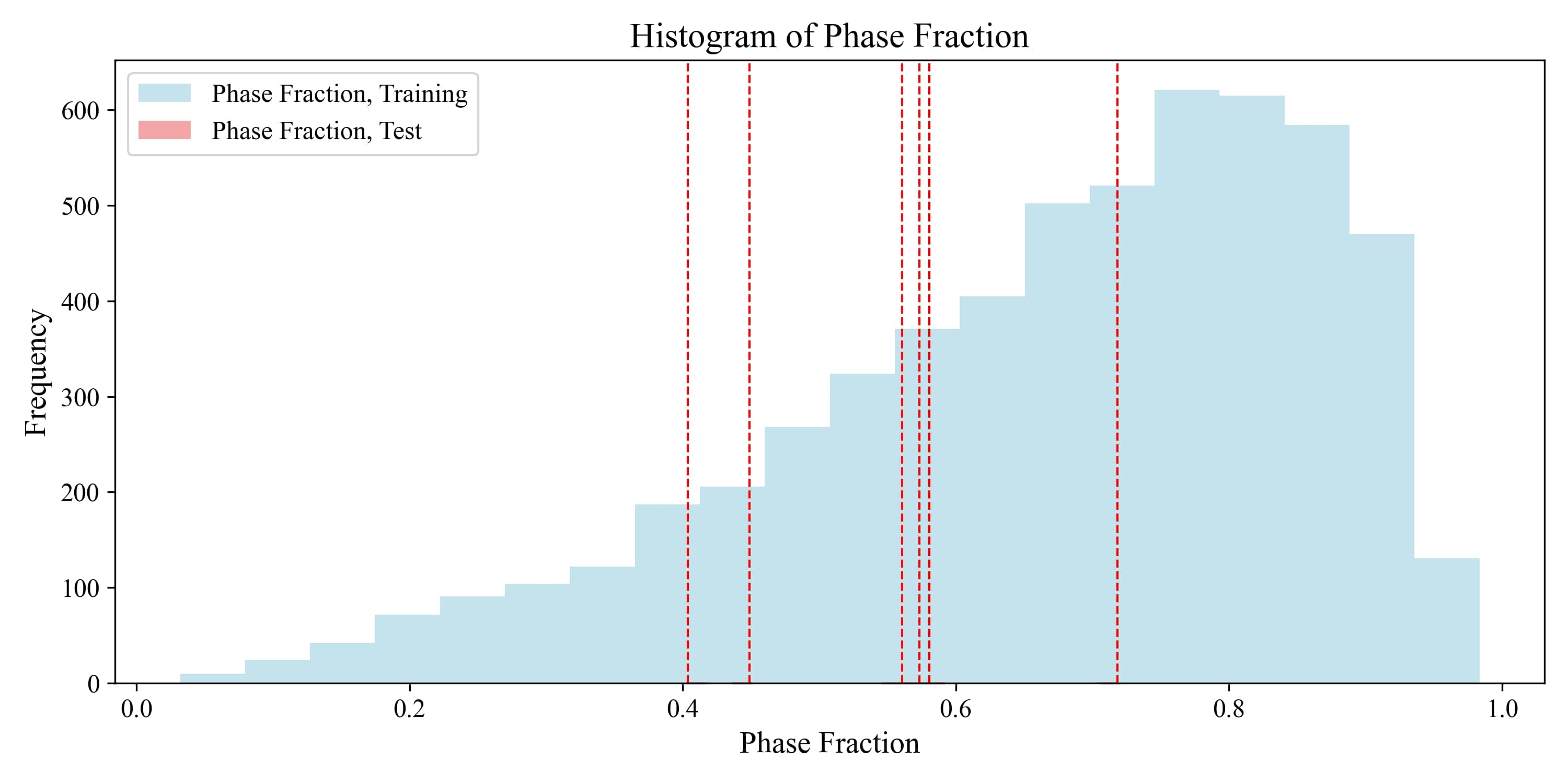}
    \caption{Phase fraction of the generated microstructures. Phase fraction refers to the proportion of each phase within a material.}
    \label{histogram_plot}
\end{figure}

Additionally, for visualization and further analysis, the phase fraction of the generated microstructures, along with the test cases, are illustrated in Fig.~\ref{histogram_plot} to showcase the diversity of the dataset. It's worth mentioning that the chosen database leans more towards higher phase fractions, which means the selected test cases are not among those with the highest frequency.

For the results in Fig.~\ref{histogram_plot}, phase fraction is calculated by determining the ratio of the volume or area occupied by the phase of interest to the total volume or area of the material. 


\section{Enhancing resolution of solution: A methodological Study}\label{sec2}

In this section, we comprehensively describe the architecture of the implemented approaches to increase the resolution of the solution map. The methods analyzed include a simple interpolation approach, a fully connected feed-forward neural network, two different versions of the MEA architecture, as the novelty of this paper, and the standard U-Net architecture. The motivation behind this study is to compare the performance of various approaches against our developed algorithm. It should be noted that the codes for all architectures have been developed in PyTorch.
\subsection{Interpolation approach}\label{interpol}
The simplest and most direct method for rescaling a solution field to a higher resolution involves interpolating the solutions using a selected order, such as linear, quadratic, or cubic. This process mirrors the method used in the finite element method, which utilizes the capabilities of shape functions. More specifically, we utilized the \texttt{ndimage.zoom} command from the \texttt{SciPy} library to resize an n-dimensional array. The \texttt{order} parameter within this command determines the interpolation method applied during the resizing process, effectively interpolating the values between existing data points to fill gaps when the array is enlarged. This interpolation can lead to smoothing, as the chosen method (specified by the \texttt{order} parameter) introduces smoother transitions between pixel values. For instance, higher-order interpolation methods, such as cubic, typically produce smoother results compared to lower-order methods like linear. This aspect of interpolation is comparable to the post-processing phase in the finite element method, where different shape functions of varying orders are employed to evaluate and interpolate solutions across the element domain based on the nodal solution values.

\begin{figure}[t]
\center
    \includegraphics[width=0.7\textwidth]{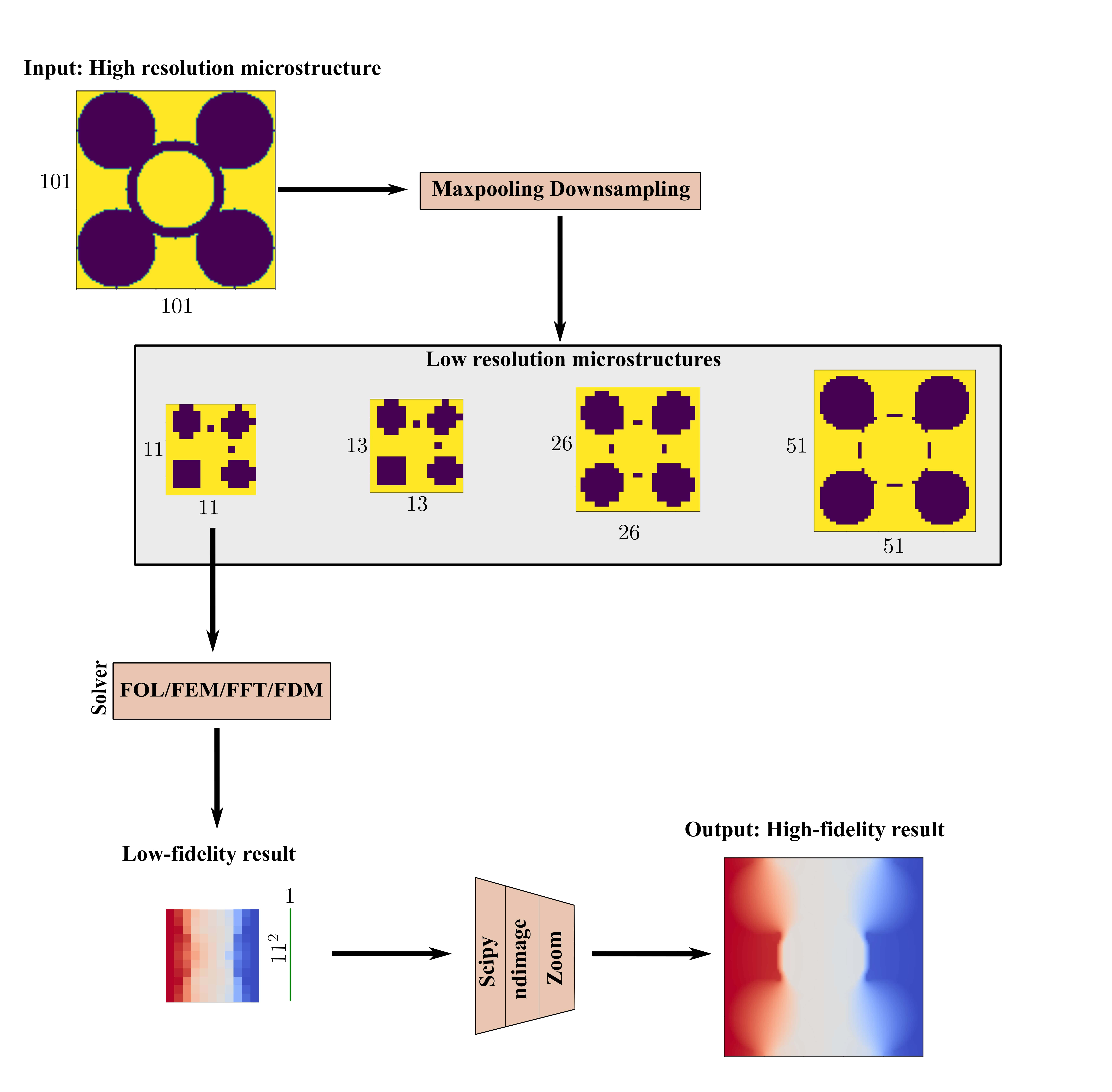}
    \caption{Implemented interpolation architecture to obtain a high-fidelity solution map from a low-fidelity one. After condensing the high-resolution parametric space into lower resolutions, the low-fidelity solution is obtained by solving the boundary value problem on the coarsest grid using FEM/FOL solver. This low-fidelity solution is then upscaled using an interpolation function to achieve higher fidelity.}
    \label{interpolation_approach}
\end{figure}

Fig.~\ref{interpolation_approach} illustrates the architecture of the interpolation approach used in this study. The input consists of the high-resolution spatial distribution of each coefficient, which is then condensed to lower resolutions \(N_{\text{51R}} = 51\), \(N_{\text{26R}} = 26\), \(N_{\text{13R}} = 13\), and \(N_{\text{11R}} = 11\) using the \texttt{Max Pooling} function. In this work, By adjusting the \texttt{pool-window-size} parameter, the original $101 \times 101$ heat conductivity maps are condensed into four different lower-resolution versions. In the next step, the lowest resolution, \(N_{\text{11R}} = 11\), is then fed into a numerical solver which potentially can be methodologies such as Finite Element Method (FEM), Finite Difference Method (FDM), or Fast Fourier Transform (FFT), to generate corresponding low-resolution FE results. In this work physics operator learning (FOL) is used to capture the low-fidelity solution. Afterward, this solution is upscaled using the \texttt{SciPy.ndimage.zoom} function.

\subsection{Feedforward neural network approach}
This method is based on a fully connected Feed Forward Neural Network (FFNN).
Each neural network follows the standard structure, featuring a single input layer, potentially several hidden layers, and an output layer. Every layer is interconnected, transferring information to the next layer \cite{schmidhuber2015deep}. Within each layer, the neurons are not interconnected. Thus, we denote the information transfer from the $l-1$ layer to $l$ using the vector $\bm{z}^l$. Each component of vector $\bm{z}^l$ is computed by
\begin{equation}
\label{eq:NN_1}
    {z}^l_m = {a} (\sum_{n=1}^{N_l} w^l_{mn} {z}_n^{l-1} + b^l_{m} ),~~~l=1,\ldots,L. 
\end{equation}
In Eq.\,(\ref{eq:NN_1}), ${z}^{l-1}_n$, is the $n$-th neuron within the $l-1$-th layer. The component $w_{mn}$ shows the connection weight between the $n$-th neuron of the layer $l-1$ and the $m$-th neuron of the layer $l$. Every neuron in the $l$-th hidden layer owns a bias variable $b_m^l$. The number $N_I$ corresponds to the number of neurons in the $l$-th hidden layer. The total number of hidden layers is $L$. The letter $a$ stands for the activation function in each neuron. The activation function $a(\cdot)$ is usually non-linear. 

\begin{figure}[t]
\center
    \includegraphics[width=\textwidth]{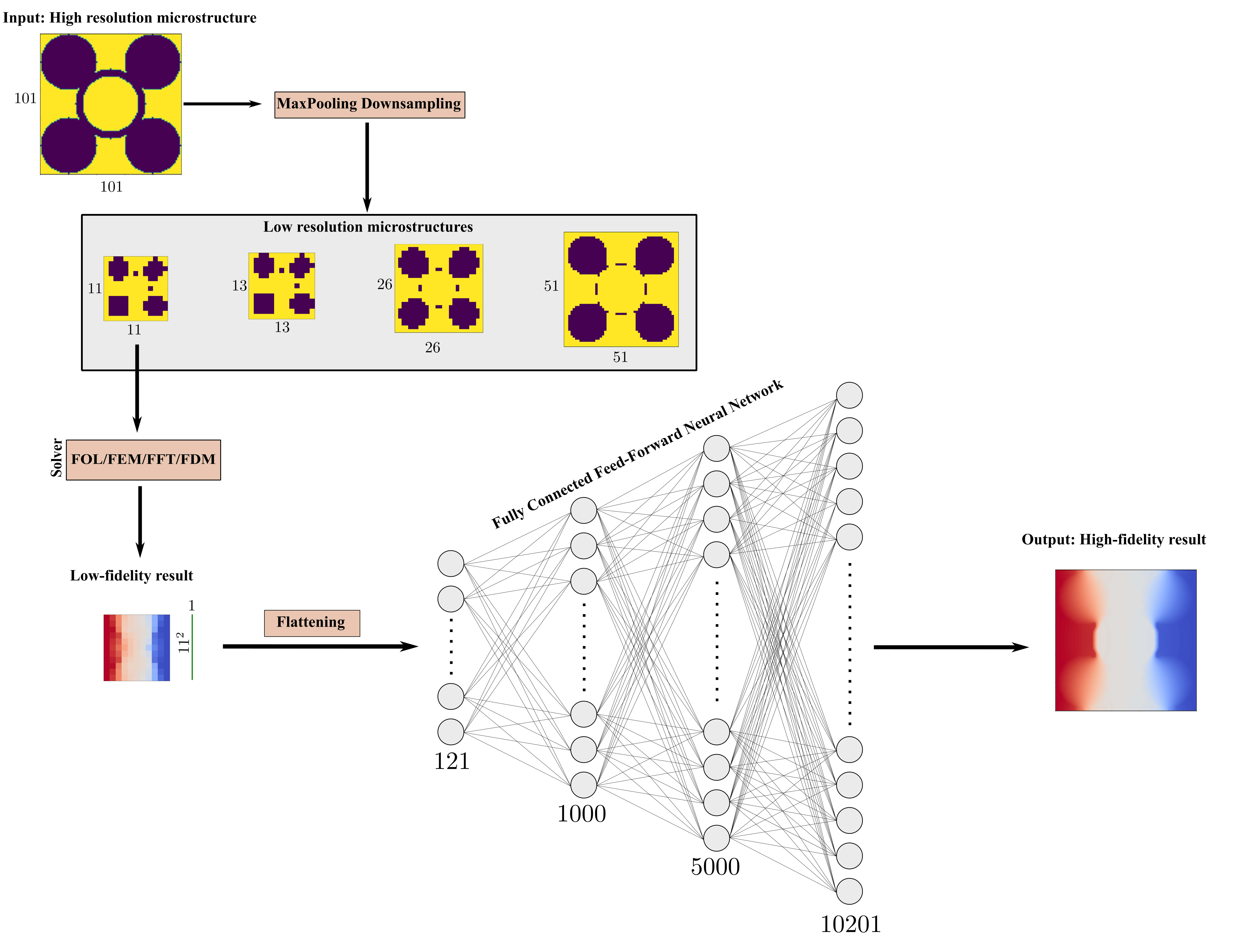}
    \caption{Fully connected feed-forward neural network (FFNN) architecture to obtain a high-fidelity solution from a low-fidelity one. Similar to the interpolation approach, the low-fidelity solution is first obtained using an FEM/FOL solver. In the FFNN architecture, however, the low-fidelity solution is first flattened and then fed into a fully connected neural network. The output of the FFNN is designed to yield the high-fidelity solution.}
    \label{FFNN_approach}
\end{figure}

The principle idea behind the implemented FFNN is depicted in Fig.~\ref{FFNN_approach}. Initially, similar to the interpolation approach prescribed in section.~\ref{interpol}, a high-resolution microstructural representation, represented by a $101 \times 101$ grid, undergoes dimensionality condensation via the \texttt{Max Pooling} algorithm, yielding a series of lower-resolution counterparts. After resolution degradation, the microstructure with a resolution of $11 \times 11$ is subjected to the FOL solver.

In the FFNN approach, a critical preprocessing step involves the transformation of the two-dimensional low-fidelity solution into a one-dimensional array, commonly referred to as flattening. This is a prerequisite for input compatibility with the FFNN architecture. The FFNN in this study is structured with an input layer succeeded by hidden layers, comprising 121, 1000, and 5000 neurons, respectively, resulting in an output layer with 10201 neurons. Within this approach, the low-fidelity solution is converted to the high-fidelity one. We utilized the Swish activation function, with a learning rate set to 1e-4. We utilized a batch size of 100 and employed the Adam optimizer for training.

\subsection{Standard U-Net}\label{U-Net Section}

\begin{figure}[t]
\center
    \includegraphics[width=\textwidth]{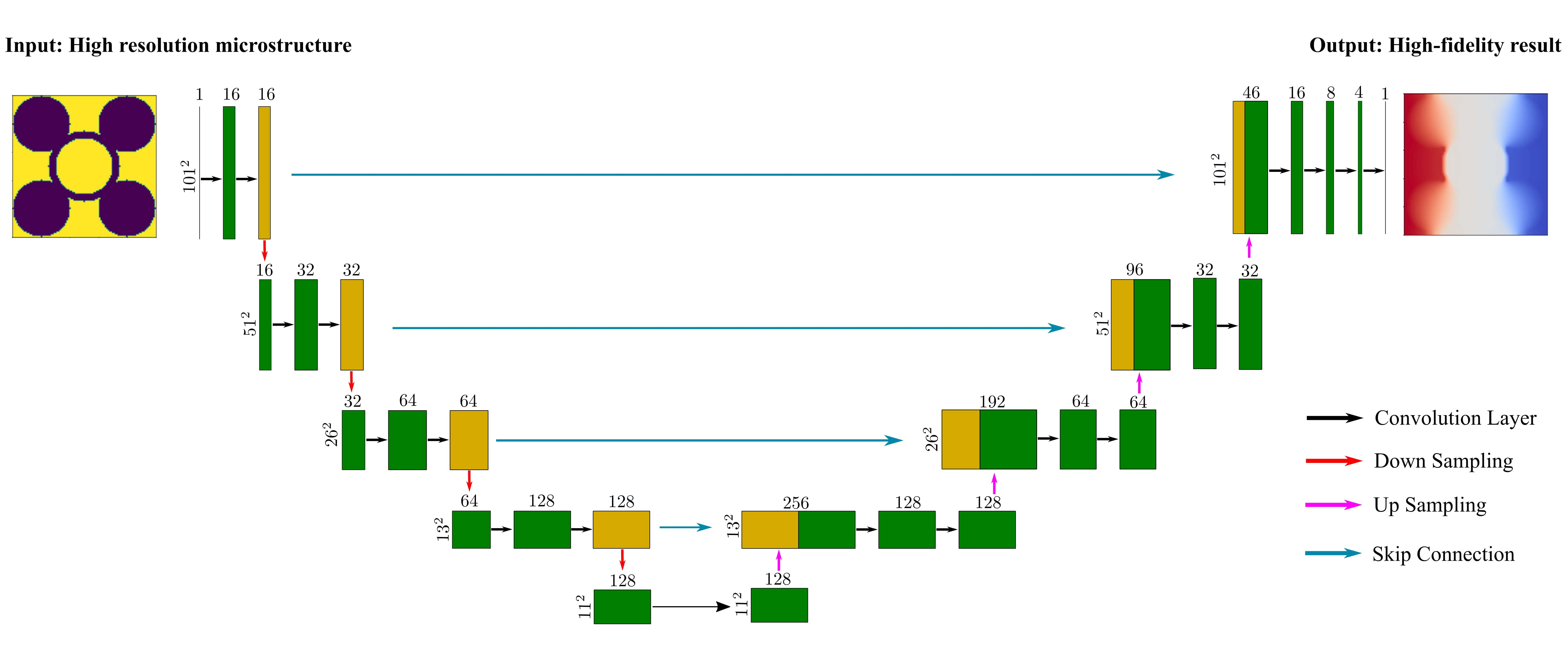}
    \caption{The classical U-Net architecture employed in the current study. The input to the U-Net is a high-resolution heat conductivity coefficient map, and the output is a high-resolution solution. Within the U-Net's decoder component, the input undergoes downsampling through four stages, resulting in a $10 \times 10$ grid featuring 11 nodes in both horizontal and vertical directions, alongside 128 channels. Following this, the input is upsampled in four stages, with skip connections from the corresponding layers of the decoder being applied at each step to enhance information flow and preserve spatial details.}
    \label{U-Net-Classic}
\end{figure}

In this study, we utilize a classic U-Net architecture to predict a high-fidelity solution map. The network architecture, illustrated in Fig.~\ref{U-Net-Classic}, diverges from the traditional U-Net application where images are the typical input; here, the spatial distribution of heat conductivity coefficients serves as the input. The U-Net structure in this work processes an input of a $1 \times 101 \times 101$ high-resolution heat conductivity map of the microstructure, generating a high-fidelity temperature distribution as the output. 

In the architecture of the standard U-Net, the encoder is designed to sequentially compress the input data while extracting critical features. It starts with a $1 \times 101 \times 101$ input representing the high-resolution thermal conductivity coefficients of the microstructure. The encoder consists of twelve convolutional layers grouped in blocks, each designed to progressively reduce the spatial dimensions and increase the channel depth to encode the input data into a compact, feature-rich representation. The convolutional layers are characterized by a kernel size of 3, with strides of 1 or 2 and padding 1 to maintain or reduce dimensions as needed. Each convolution operation is followed by batch normalization and ReLU activation functions that ensure efficient feature extraction and introduction of non-linearity. The design of the encoder facilitates the acquisition of hierarchical features from the spatial heat conductivity map, reducing the final output dimension to $128 \times 11 \times 11$, ready for the decoder phase.

The decoder part of the U-Net architecture inversely mirrors the encoder and aims to reconstruct the high-resolution output from the encoded features. It utilizes a series of convolutional layers and up-convolutional layers along with skip connections that integrate corresponding encoder features to enrich the decoder input at various stages. Starting from a compact $128 \times 11 \times 11$ representation, the decoder gradually increases the spatial dimensions while refining the feature maps to reconstruct the final output. This is achieved by alternating convolutional layers that increase feature resolution and upsample layers that expand spatial dimensions, aided by skip connections that reintegrate spatial details from the encoder. The upsampling process ends in a final convolution layer that produces the high-fidelity solution with dimensions that match the original input, effectively capturing the detailed temperature distribution within the microstructure. 


\subsection{Microstructure-embedded autoencoder (MEA)}

Similar to the previous approaches, a crucial step in MEA architecture involves creating lower-resolution representations of heat conductivity maps. To achieve this, once again we employed the \texttt{Max Pooling} operation by utilizing the \texttt{maximum-filter} function from the \texttt{scipy.ndimage} library in Python.

In the subsequent phase of the analysis, the lowest resolution heat conductivity map, now represented by an $11 \times 11$ matrix, is inputted into a pre-trained FOL network to solve the boundary value problem. It's important to note, however, that solving this problem isn't strictly limited to FOL; alternative solvers can also be utilized based on the specific requirements or constraints of the study. 

\begin{figure}[H]
\center
    \includegraphics[width=\textwidth]{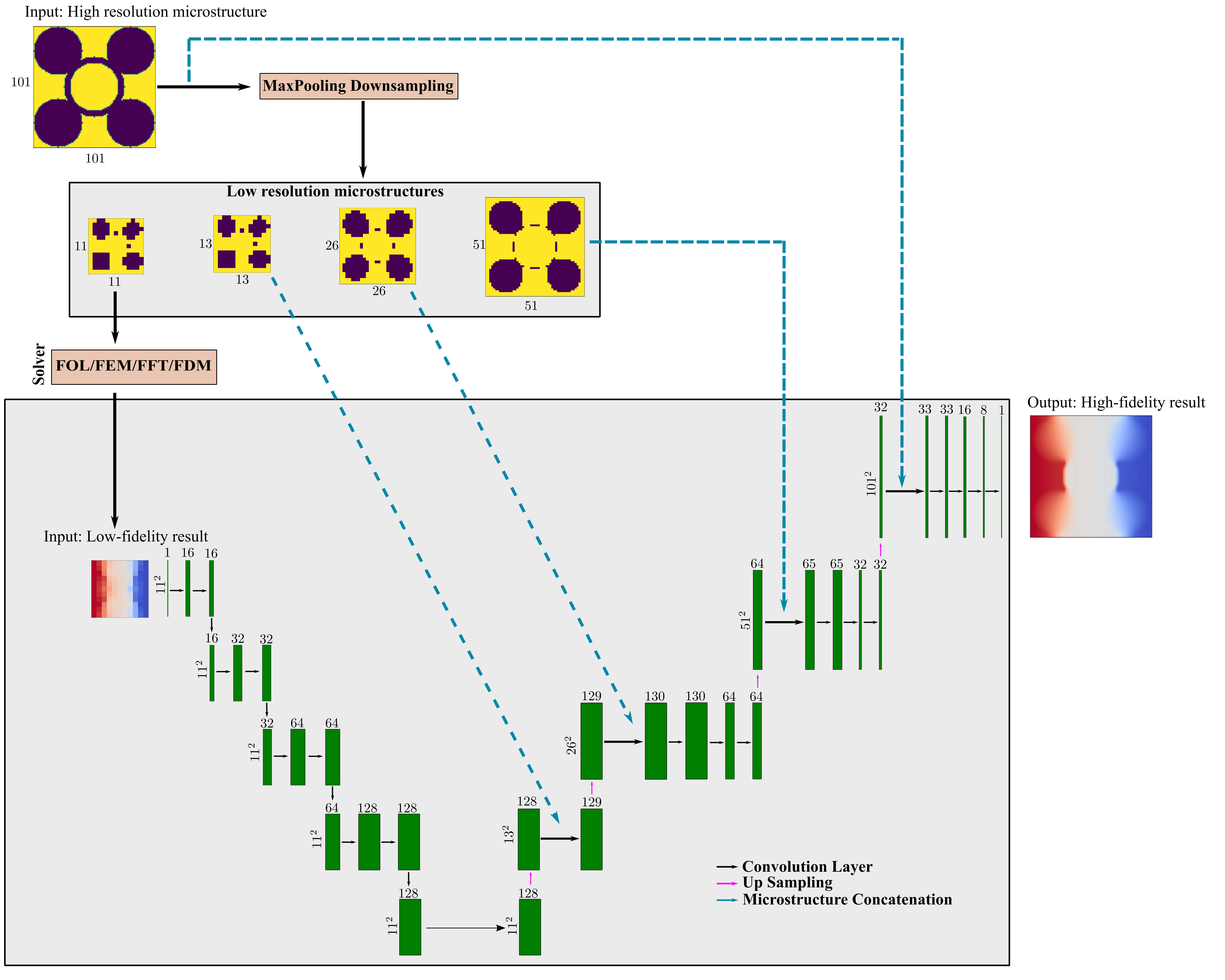}
    \caption{microstructure-embedded autoencoder architecture for a high-resolution output generation: Input is a low-resolution solution that is obtained from PINNs on an $11 \times 11$ node grid in this study and output is a high-resolution $100 \times 100$ grid with 101 nodes in the vertical and horizontal direction. The decoder integrates heat conductivity maps of sizes $13 \times 13$, $26 \times 26$, $51 \times 51$, and $101 \times 101$, concatenating them at corresponding stages of the decoder for enhanced output resolution. }
    \label{enhanced_decoder}
\end{figure}

As can be seen in Fig.~\ref{enhanced_decoder}, the encoder segment of MEA architecture starts with an input of dimensions $1 \times 11 \times 11$, which represents a low-resolution solution of a steady-state heat transfer problem. It consists of twelve convolutional layers, each with a kernel size of 3, a stride of 1, and a padding of 1. This configuration ensures that the spatial dimensions are maintained while the depth of the feature maps increases, allowing detailed feature extraction without loss of spatial information. Starting from a single channel, the encoder increases the depth to 128 channels through successive convolutional stages, each followed by batch normalization and ReLU activation to improve feature normalization and introduce nonlinearity.

The decoder part of the MEA architecture is specifically designed to reconstruct the input's high-resolution spatial features obtained from the feature representation (channels) produced by the encoder. This process involves expanding the spatial dimensions of the feature maps while integrating additional morphological information to enhance the fidelity of the reconstruction. The decoder segment begins its process with a feature map of dimensions $128 \times 11 \times 11$, which is the output of the encoder section. The decoder section encompasses four upsampling operations, systematically enlarging the spatial dimensions of the feature maps at each stage. 

\begin{figure}[t]
\center
    \includegraphics[width=\textwidth]{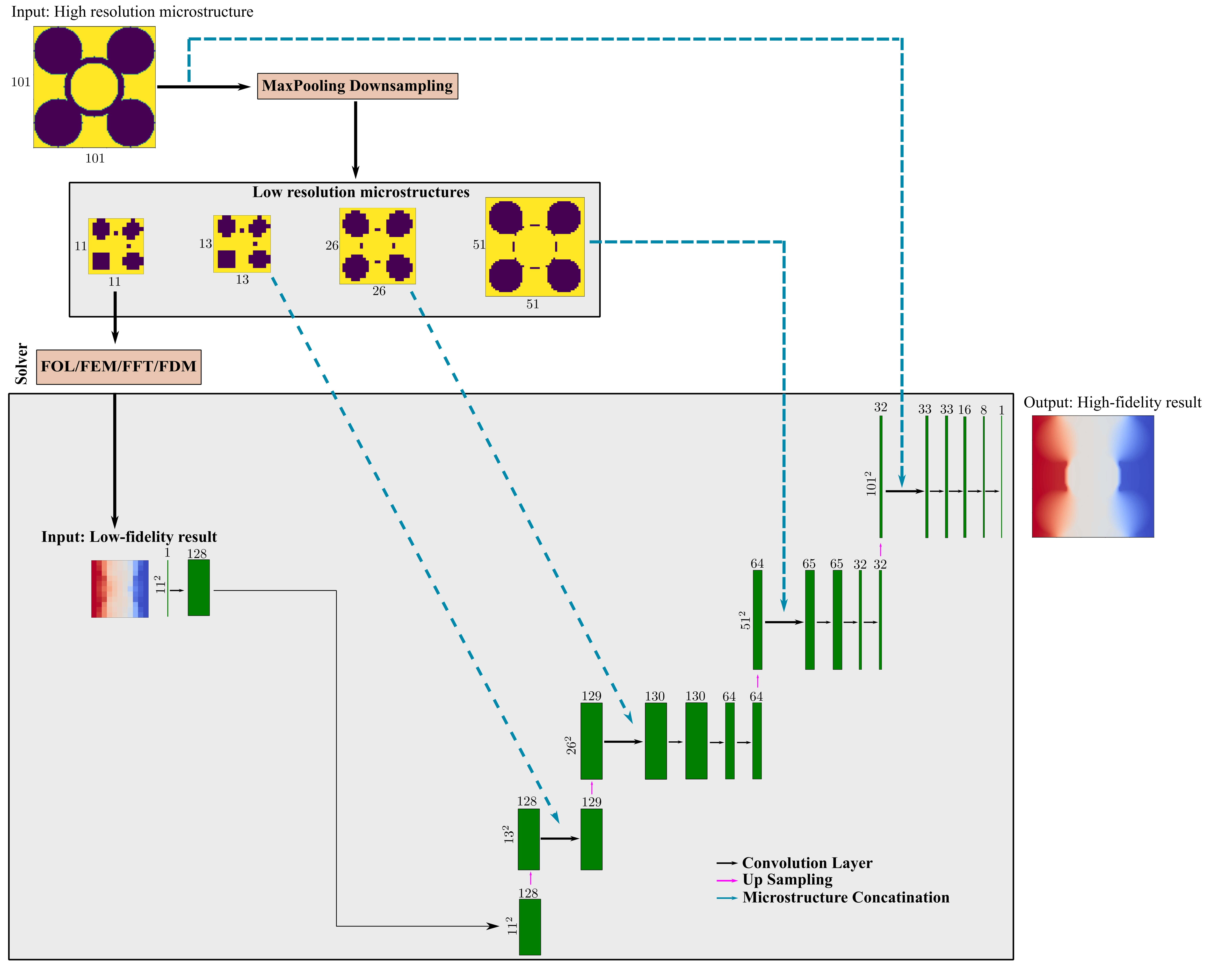}
    \caption{Second version of the microstructure-embedded autoencoder with a simplified encoder featuring a single convolution layer, maintaining the original decoder structure with microstructure concatenation as shown in Fig.\ref{enhanced_decoder}.}
    \label{single_encoder}
\end{figure}

Beginning with the channels received from the encoder, the decoder initiates its first upsampling to expand the dimensions to $13 \times 13$. This step is followed by subsequent enlargements to $26 \times 26$, then to $51 \times 51$, and ultimately achieving a spatial dimension of $101 \times 101$. After each upsampling operation, the corresponding morphological map which is the spatial distribution of the heat conductivity is integrated with the upsampled feature map in a concatenation step, so that the decoder's input is enriched with essential external spatial details that are crucial for accurate reconstruction. In general, the concatenation operation involves joining two or more matrices along a specified direction. For example, if matrix \( A_{\mathbf{m \times n}} \) and matrix \( B_{\mathbf{m \times p}} \), then concatenating these two matrices along the first dimension would result in matrix \( C_{\mathbf{m \times (n+p)}} \). Subsequent convolutional layers refine these merged feature maps by adjusting the channel depths and enhancing the spatial details to ensure that the morphological information is perfectly integrated.

This intricate decoder structure, characterized by the integration of morphological maps and a careful feature refinement process, is critical to the network's ability to reconstruct a high-fidelity solution map from the low-resolution one. Through a strategic combination of convolutional and transposed convolutional operations and the selective incorporation of external morphological data, the architecture ensures that the reconstructed output authentically reflects the high-resolution attributes of the original input. 

Considering the architecture of the modified autoencoder, in the encoder section, the low-fidelity solution undergoes twelve convolutional transformations to generate an encoded output with 128 channels. This raises a question about the necessity of utilizing 9 convolutional layers, given that a similar number of channels can be produced by a single convolutional layer in the decoder section. To address this question, an alternative architecture is presented, as shown in Fig.~\ref{single_encoder}, where only a single convolutional layer is employed to generate an output that meets the input requirements for the decoder section. In the results section, the impact of the number of convolutional layers in the decoder section on the prediction accuracy of the enhanced autoencoder is discussed.

\section{Results}\label{sec4}
In this section, we analyze the prediction outcomes of the implemented architecture to obtain the high-fidelity solution. As observed in Table~\ref{tab:methods}, the analyzed methods include the interpolation approach, FFNN, two different MEA algorithms, and a standard U-Net, which are applied across six test cases (see Fig.~\ref{test_cases}). Additionally, Table~\ref{tab:methods} also presents the methods implemented to obtain a low-fidelity solution after the condensation of the high-fidelity parametric space.

\begin{table}[h]
\centering
\caption{Methods for obtaining a high-fidelity solution map from high-fidelity microstructure data analyzed in this study.}
\label{tab:methods}
\begin{tabular*}{0.8\textwidth}{@{\extracolsep{\fill}}lc}
\toprule
Method &  Solver \\
\midrule
Interpolation approach & only low-fidelity FOL/FEM \\
Feed Forward Neural Network & low+high-fidelity FOL/FEM \\
Microstructure-embedded autoencoder, Type 1 & low+high-fidelity FOL/FEM \\
Microstructure-embedded autoencoder, Type 2  & low+high-fidelity FOL/FEM \\
Classical U-Net & only offline high-fidelity FEM \\
\botrule
\end{tabular*}
\end{table}

\subsection{Hyperparameter analysis}
In this section, we investigate the effects of two crucial hyperparameters, batch size and the extent of microstructure concatenation, on the prediction accuracy of the models. We set the learning rate for all DNNs in this study to $1 \times 10^{-4}$ and use the Adam optimizer to optimize weights and biases. The dataset is consistently split, with 80\% used for training and the remaining 20\% for validation. All DNN analyses are performed over a uniform duration of 500 epochs, with mean squared error (MSE) serving as the loss function for the analyses.

In the first study, we investigate the effects of microstructural concatenation, focusing on the integration of different resolutions. The MEA architecture shown in Fig.~\ref{enhanced_decoder} is characterized by a unique approach in the decoder part: The spatial distribution of thermal conductivity coefficients from four different resolutions is concatenated once each. This approach prompts us to investigate the effects of concatenating microstructure information from multiple resolutions. To address this question, we conducted a comprehensive study to investigate the influence of different numbers of resolution-specific concatenations.

To investigate the effects of varying concatenations, modifications were made to the decoder segment of the MEA architecture. Specifically, the approach to concatenation was adjusted to evaluate different numbers of concatenations involving spatial distributions of heat conductivity coefficients at various resolutions. In \emph{Concatenation 1}, only the distribution with a resolution of \(N_{\text{13R}} = 13\) is concatenated. \emph{Concatenation 2} involves combining the spatial heat conductivity coefficients at resolutions of \(N_{\text{13R}} = 13\) and \(N_{\text{26R}} = 26\). For \emph{Concatenation 3}, resolutions of \(N_{\text{13R}} = 13\), \(N_{\text{26R}} = 26\), and \(N_{\text{51R}} = 51\) are concatenated, and finally, in \emph{Concatenation 4}, heat conductivity coefficients from the resolutions of \(N_{\text{13R}} = 13\), \(N_{\text{26R}} = 26\), \(N_{\text{51R}} = 51\), and \(N_{\text{101R}} = 101\) are concatenated together. A batch size of 50 is used in this study.

\begin{figure}[t]
\center
    \includegraphics[width=0.6\textwidth]{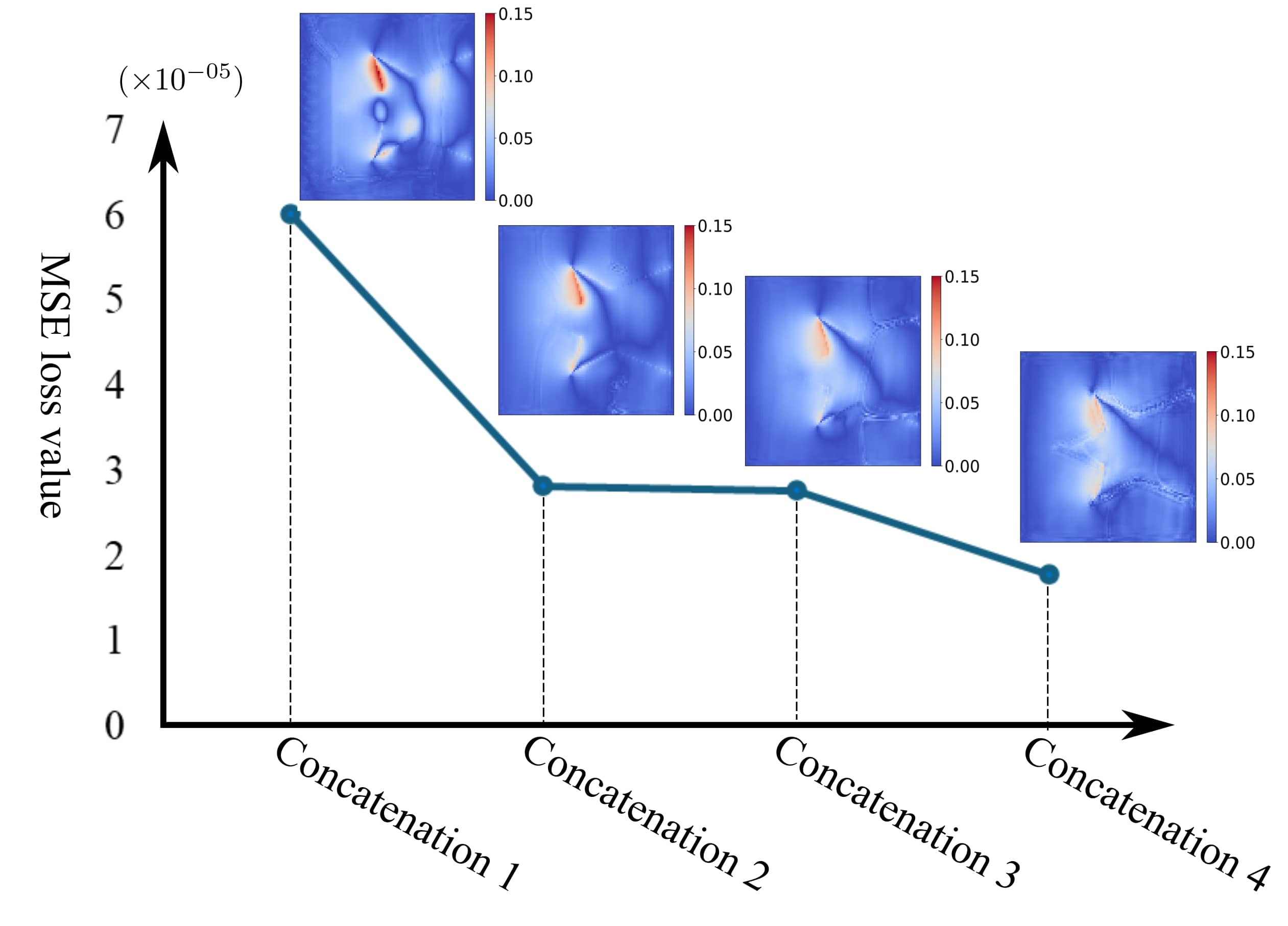}
    \caption{Exploration of various concatenation strategies within the decoder segment of the enhanced autoencoder. Each panel represents a distinct scenario of concatenation: Concatenation 1 illustrates the model utilizing a singular $13 \times 13$ resolution of concatenation. Moving to Concatenation 2, both $13 \times 13$ and $26 \times 26$ resolutions are concatenated. Concatenation 3 broadens the scope by including $13 \times 13$, $26 \times 26$, and $51 \times 51$ resolutions. The most comprehensive strategy, Concatenation 4, applies data from $13 \times 13$, $26 \times 26$, $51 \times 51$, and $101 \times 101$ resolutions. The color maps across the panels highlight the absolute error between the network's predictions and the ground truth calculations.}
    \label{concat_study}
\end{figure}

Figure \ref{concat_study} presents the results from analyzing the impact of the number of microstructural concatenations. It is evident from the error graph that an increase in the number of concatenations correlates with a reduction in error. Given that in the process of microstructural concatenation, only a single channel is added to the corresponding stage of the decoder section, this approach does not significantly affect computational cost. Therefore, for all subsequent studies, the \emph{Concatenation 4} method is employed. 

\begin{figure}[t]
\center
    \includegraphics[width=0.6\textwidth]{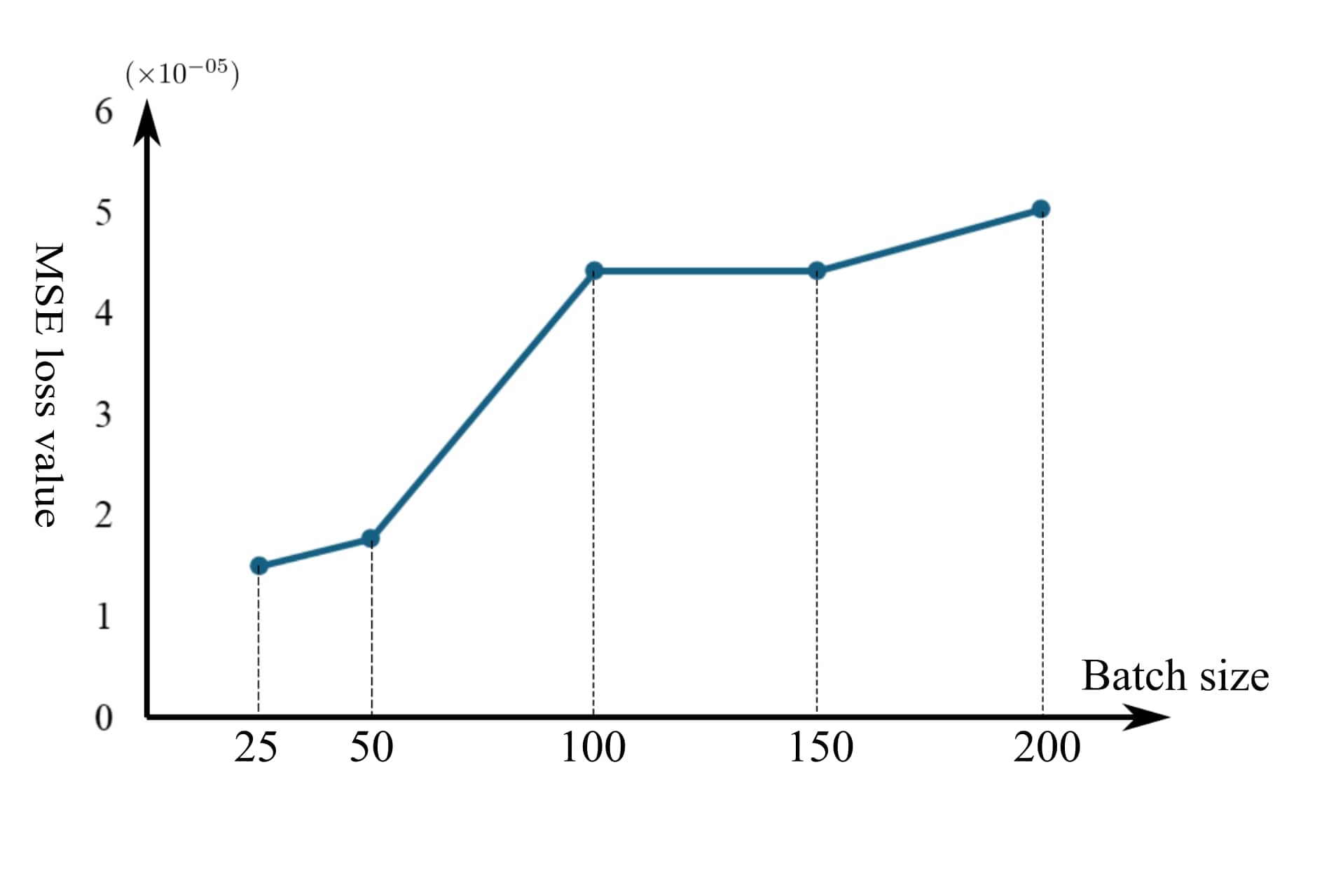}
    \caption{Investigation of the effects of varying batch sizes on DNNs performance, under a uniform experimental setup employing the Concatenation 4 technique. Throughout this study, the learning rate and optimizer settings remain constant.}
    \label{batch_size_study}
\end{figure}

The other hyperparameter analyzed is the batch size, a crucial factor in deep learning, which refers to the number of samples processed in a single forward and backward pass of the network. It establishes a balance between accuracy, efficiency, and speed. Larger batch sizes can speed up training but can reduce accuracy and increase the risk of overfitting. Conversely, smaller stacks improve accuracy by acting as regularizers and reducing generalization error, although they require more computational resources and increase training time. In this study, we investigate the influence of four different batch sizes on model performance, implementing the \emph{Concatenation 4} technique as a uniform experimental condition to ensure consistency.

Figure \ref{batch_size_study} depicts the impact of varying batch sizes on the model's error rate. It is observed that reducing the batch size leads to a decrease in error, highlighting a direct relationship between batch size and model accuracy. However, this reduction in batch size comes at the cost of increased computational resources. After considering the balance between error rates and computational demand, a batch size of 50 is identified as the optimal choice for further investigations. This batch size effectively balances the trade-off between achieving high accuracy and managing computational costs.

\subsection{Quality of high-fidelity solutions from studied architectures}
\begin{figure}[t]
    \centering
    \raisebox{1ex}{}\hspace{0.14\textwidth}%
    \raisebox{1ex}{Prediction}\hspace{0.06\textwidth}%
    \raisebox{1ex}{FE Results}\hspace{0.03\textwidth}%
    \raisebox{1ex}{Absolute Error}
    \begin{subfigure}{0.65\textwidth}
        \includegraphics[width=\textwidth]{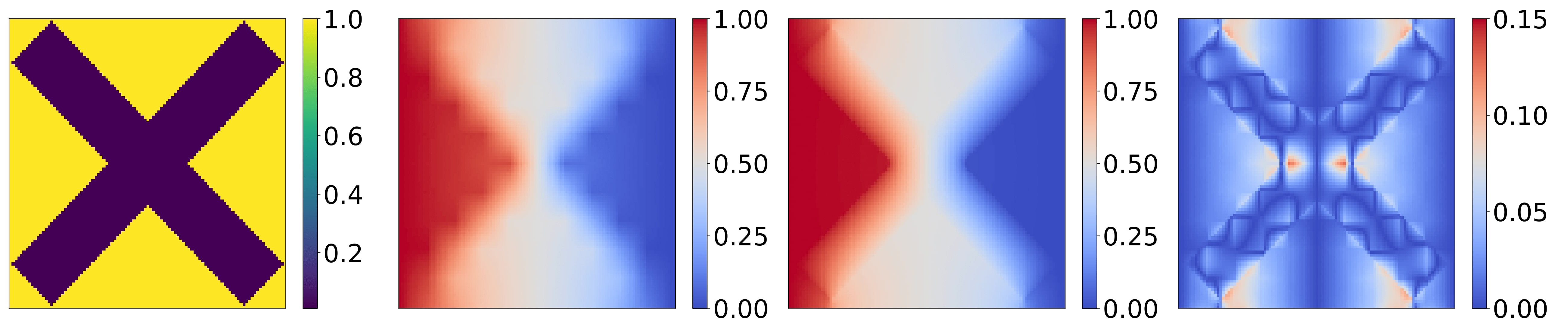}
        \end{subfigure}
    
    \begin{subfigure}{0.65\textwidth}
        \includegraphics[width=\textwidth]{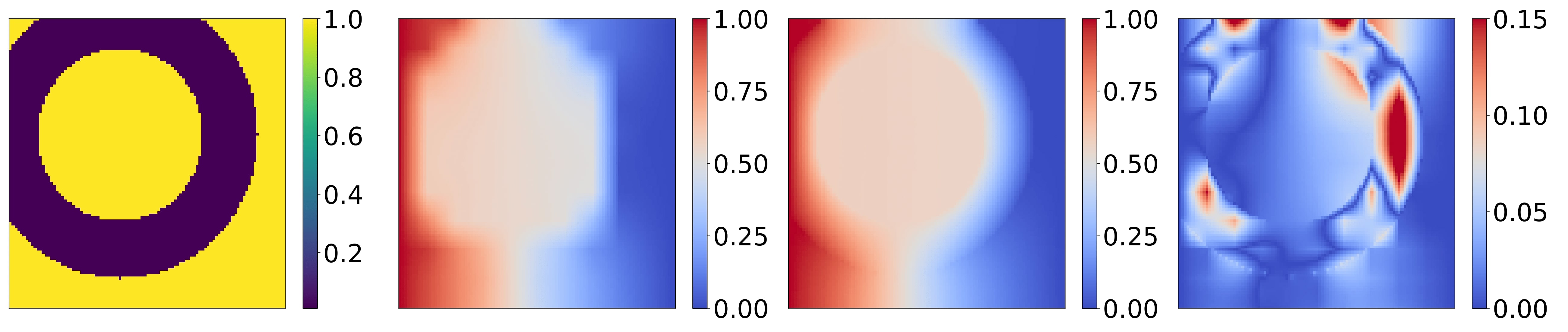}
    \end{subfigure}
    
    \begin{subfigure}{0.65\textwidth}
        \includegraphics[width=\textwidth]{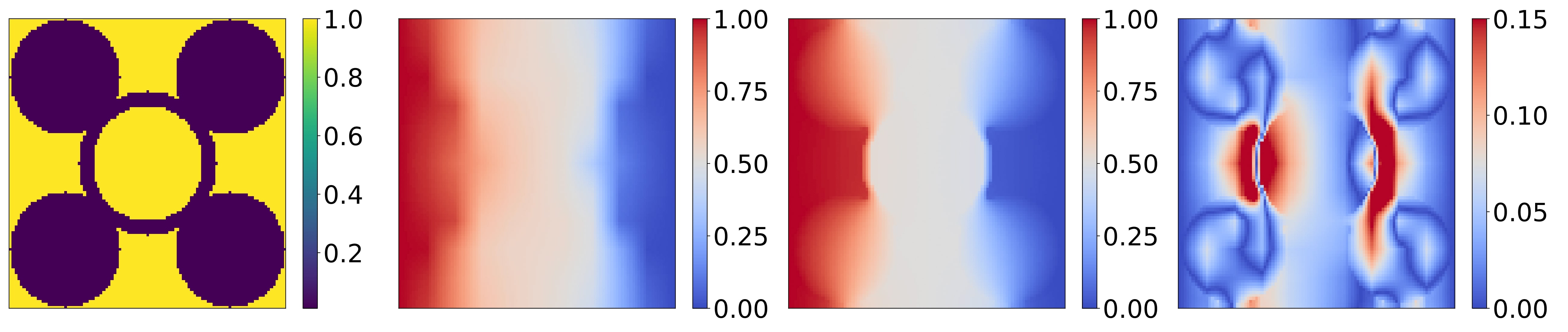}
    \end{subfigure}   
    \caption{Prediction results of thermal diffusion using the interpolation approach across three test scenarios. The contour plot in this and other similar figures shows the spatial distribution of the temperature.}
    \label{interpolation}
\end{figure}

\begin{figure}[t]
    \raisebox{1ex}{}\hspace{0.14\textwidth}%
    \raisebox{1ex}{Prediction}\hspace{0.06\textwidth}%
    \raisebox{1ex}{FE Results}\hspace{0.03\textwidth}%
    \raisebox{1ex}{Absolute Error}
    \centering    
    \begin{subfigure}{0.65\textwidth}
        \includegraphics[width=\textwidth]{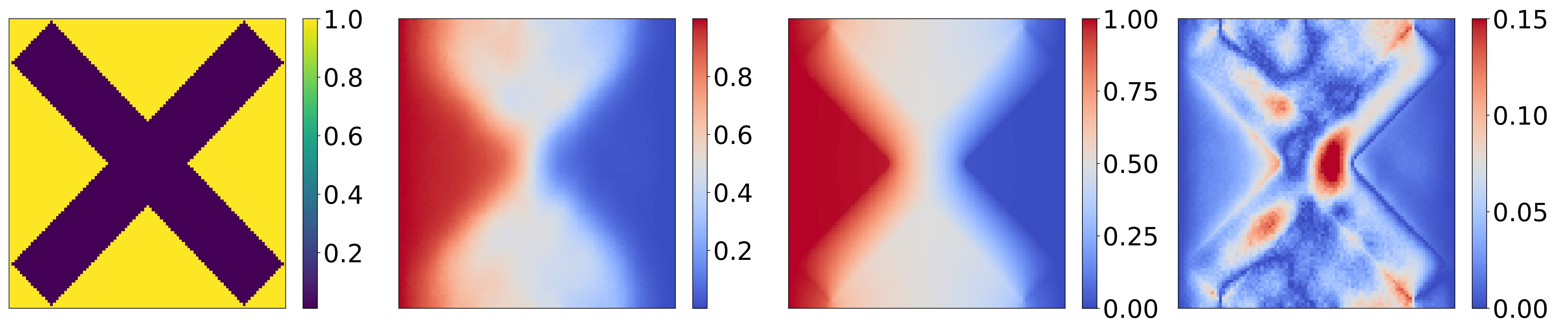}
        \end{subfigure}
    
    \begin{subfigure}{0.65\textwidth}
        \includegraphics[width=\textwidth]{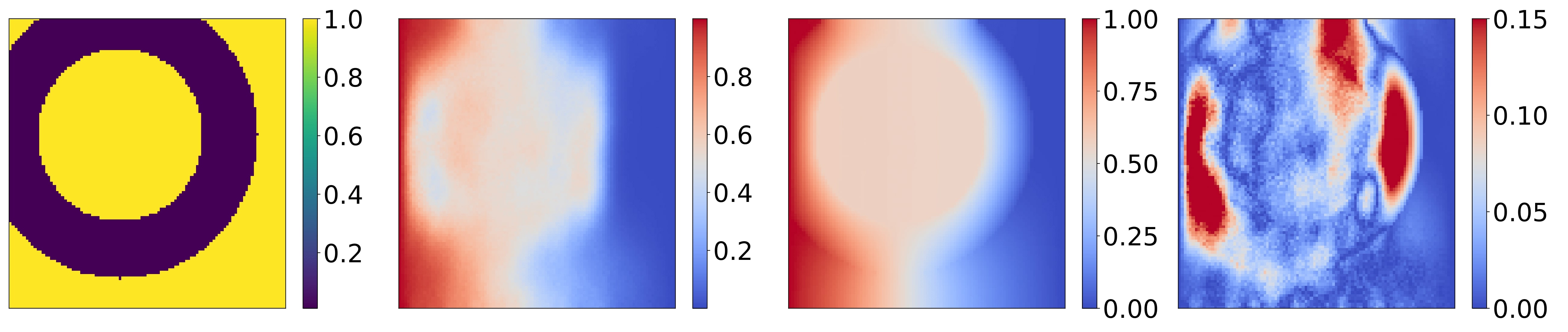}
    \end{subfigure}
    
    \begin{subfigure}{0.65\textwidth}
        \includegraphics[width=\textwidth]{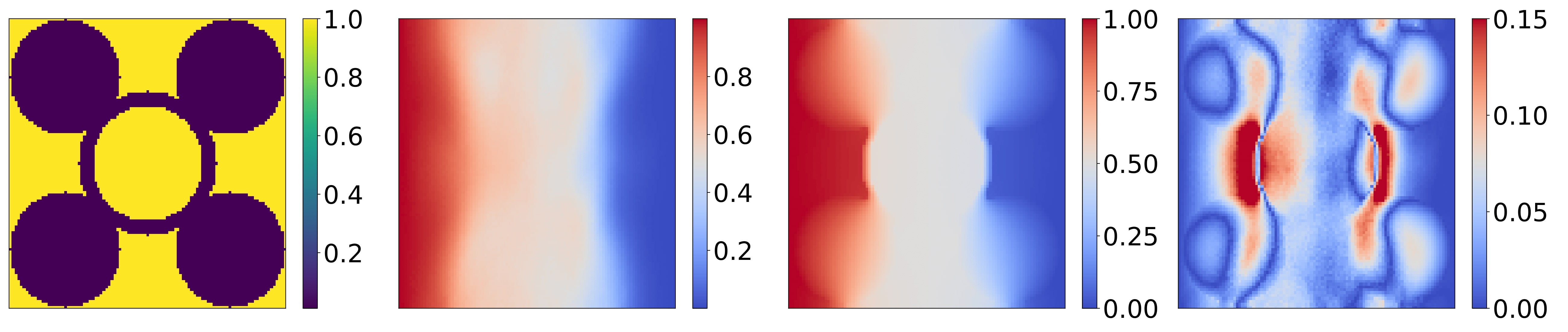}
    \end{subfigure}   
    \caption{Prediction results of thermal diffusion using the FFNN across three test scenarios.}
    \label{FFNN}
\end{figure}

Following a comprehensive hyperparameter analysis with a focus on batch size and concatenation strategies, we proceeded to assess the performance of the utilized architectures on out-of-sample test cases, as illustrated in Fig.~\ref{test_cases}. Beyond examining the direct outputs of these DNNs, which represent 2D temperature distributions, we deepen our comparison by computing the gradients of the DNN outputs. These gradients were then multiplied with spatial heat conductivity coefficient maps to calculate the heat flux. This additional step was crucial for a thorough evaluation, enabling us to determine whether the DNNs used in this study could accurately capture gradient terms. The results of the heat flux calculations are summarized in Appendix A. 

\begin{figure}[t]
\center
    \includegraphics[width=\textwidth]{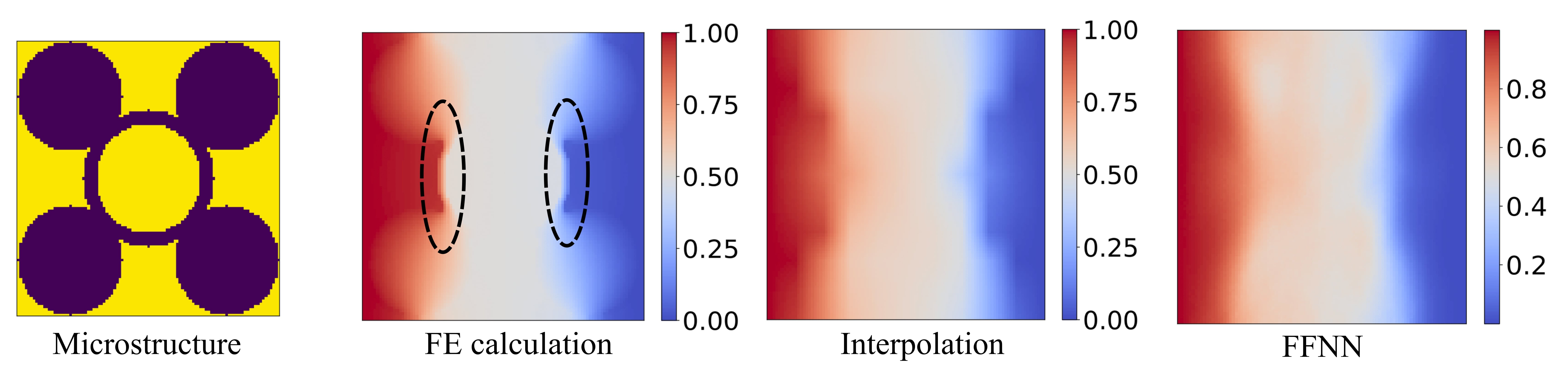}
    \caption{A comparative analysis of interpolation and FFNN upscaling methods applied to a two-dimensional microstructure model. The image labeled FE calculation shows the result of high-fidelity finite element analysis, in which sharp interfaces within the microstructure are highlighted, as indicated by the dashed ellipses. The Interpolation and FFNN images show the upscaled results of the low-fidelity solution of the microstructure under consideration. As can be seen, compared to the high-fidelity FE calculation, the remarkable discrepancies in detail preservation, especially at sharp interfaces, show the limitations of the interpolation and FFNN approaches.}
    \label{sharp_interface}
\end{figure}

Fig.~\ref{interpolation} and Fig.~\ref{FFNN} present the prediction results from the interpolation and FFNN architectures, respectively. As shown in the figures, both the interpolation and feedforward neural network techniques frequently fail to accurately capture detailed edge patterns. This case is clearly visible in Fig.~\ref{sharp_interface}. This discrepancy is due to several factors. Interpolation methods, including linear, quadratic, and cubic approaches, rely on mathematical formulas to interpolate new data points from a known set of points. Although useful, these methods are basic and localized. They may not fully capture the complexities of patterns that are highly variable or non-linear, often smoothing out results and failing to represent sharp edges and complex details. On the other hand, feedforward neural networks also confront a unique set of challenges. By processing input data as a one-dimensional array (in flattening operation), FFNNs lose critical spatial information that is essential for detecting complex patterns, such as edges. The lack of convolutional or pooling operations further reduces FFNNs' capabilities in identifying and reconstructing significant high-level features necessary for high-resolution output. 

In contrast, architectures such as MEA, which contain convolutional layers similar to the U-Net model, are characterized not only by the recognition of spatial hierarchies and dependencies but also by the reinforcement of the model through the concatenation of microstructures of various resolutions. The convolutional layers in MEA can use special filters to recognize edges and textures, regardless of their position in the input. In addition, the concatenation of microstructures helps to preserve detailed information, preventing the loss of high-resolution detail that typically occurs during downsampling.

\begin{figure}[t]
    \raisebox{1ex}{}\hspace{0.14\textwidth}%
    \raisebox{1ex}{Prediction}\hspace{0.06\textwidth}%
    \raisebox{1ex}{FE Results}\hspace{0.03\textwidth}%
    \raisebox{1ex}{Absolute Error}
    \centering   
    \begin{subfigure}{0.65\textwidth}
        \includegraphics[width=\textwidth]{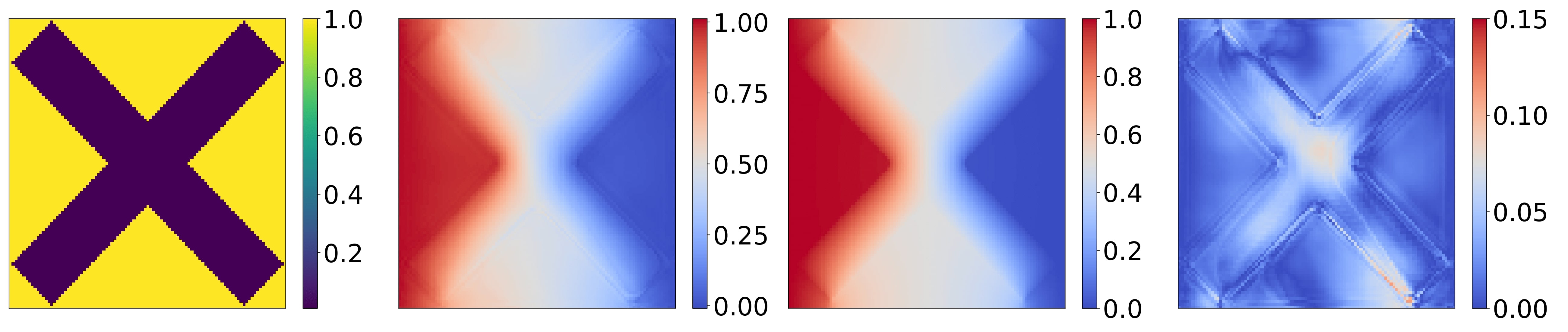}
        \end{subfigure}
    \begin{subfigure}{0.65\textwidth}
        \includegraphics[width=\textwidth]{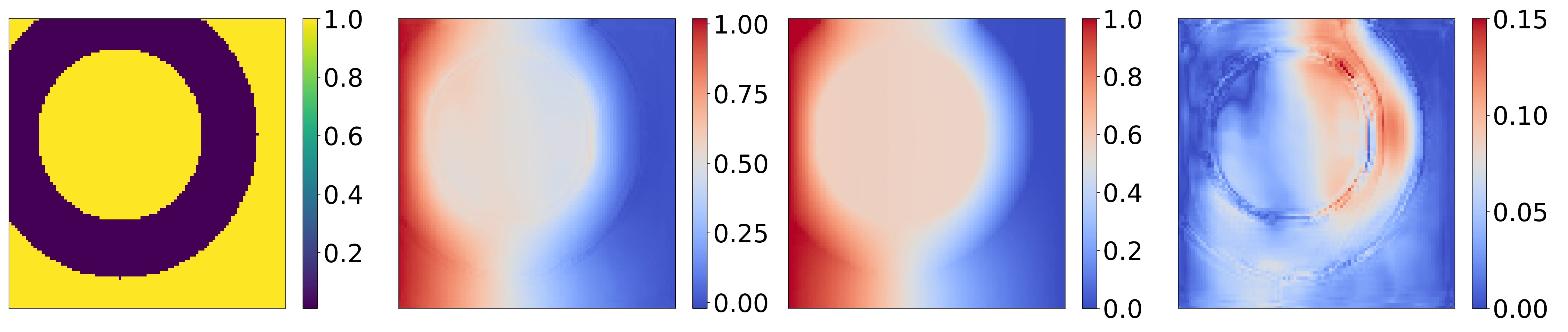}
    \end{subfigure}   
    \begin{subfigure}{0.65\textwidth}
        \includegraphics[width=\textwidth]{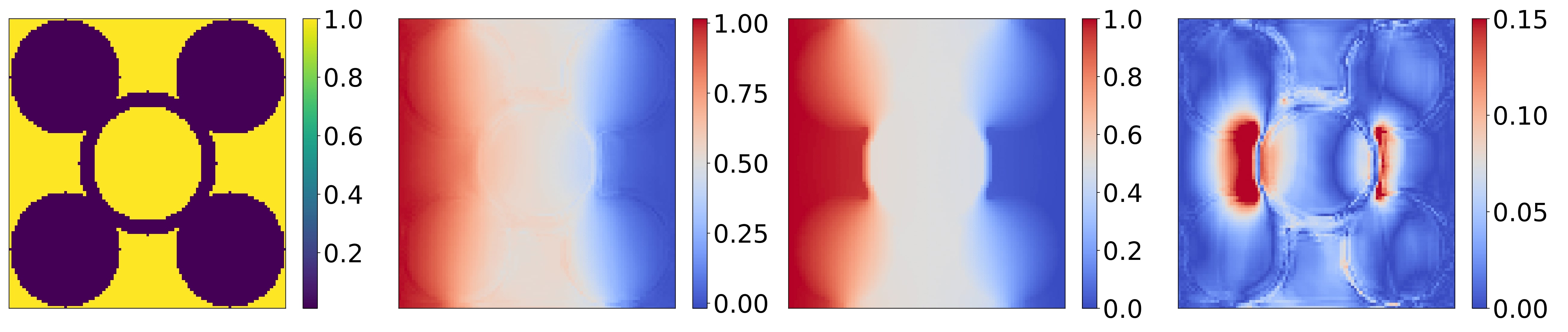}
    \end{subfigure}
    
    \caption{Prediction results of the thermal diffusion using the MEA-Type 1 architecture, which incorporates twelve convolutional layers in the encoder section are demonstrated in three test cases.}
    \label{9_layer_autoencoder}
\end{figure}

Fig.~\ref{9_layer_autoencoder} showcases the predictive capabilities of the MEA architecture in comparison to the FE solutions for the heat conductivity problem. The error column in Fig.~\ref{9_layer_autoencoder} reveals surprisingly effective performance by the network on severely OOD test cases. This accuracy is evident in the absolute error color map, highlighting the robustness and adaptability of the MEA architecture in accurately modeling complex patterns even when confronted with data that significantly deviates from the distribution observed during training. This capability originates from the use of convolutional layers, which not only recognize spatial data but also enhance the model by concatenating microstructures throughout the decoder segment. These layers use specialized filters to detect edges and textures regardless of their position in the input, while the concatenation process enhances the decoder segment to preserve highly detailed information during upsampling.

\begin{figure}[t]
    \raisebox{1ex}{}\hspace{0.14\textwidth}%
    \raisebox{1ex}{Prediction}\hspace{0.06\textwidth}%
    \raisebox{1ex}{FE Results}\hspace{0.03\textwidth}%
    \raisebox{1ex}{Absolute Error}
    \centering   
    \begin{subfigure}{0.65\textwidth}
        \includegraphics[width=\textwidth]{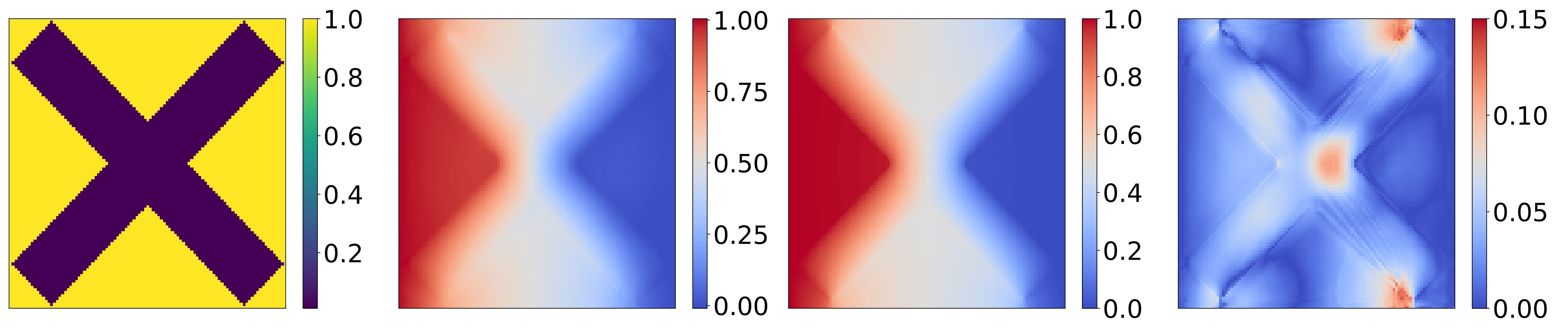}
        \end{subfigure}
    \begin{subfigure}{0.65\textwidth}
        \includegraphics[width=\textwidth]{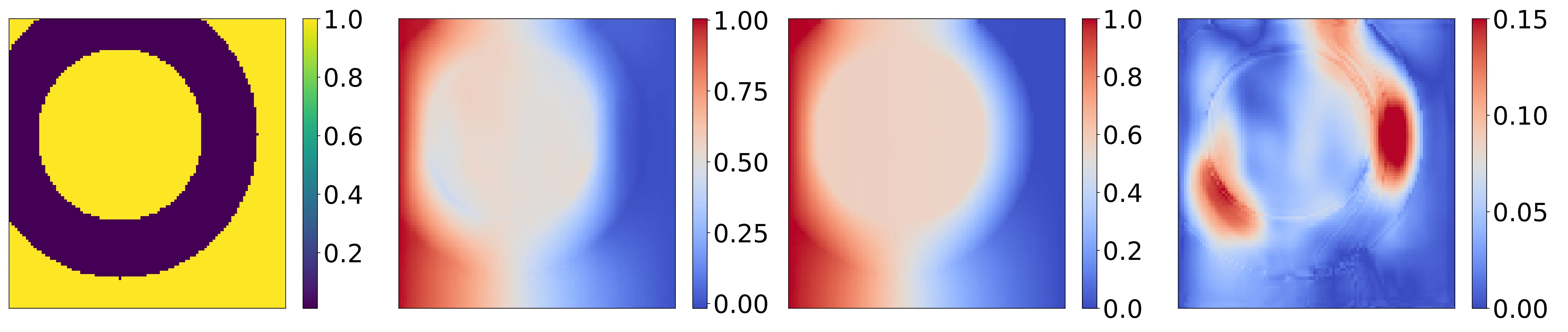}
    \end{subfigure}   
    \begin{subfigure}{0.65\textwidth}
        \includegraphics[width=\textwidth]{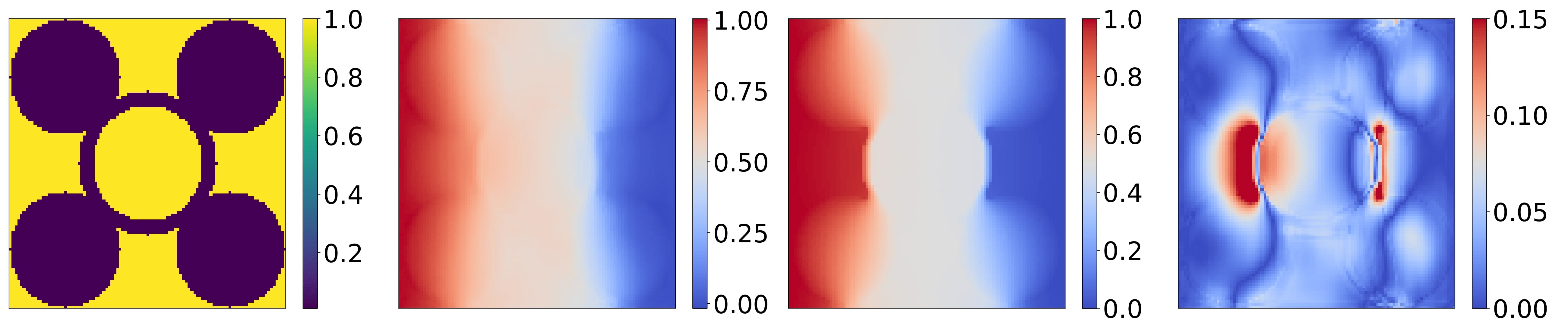}
    \end{subfigure}
    
    \caption{Prediction results of the thermal diffusion using the MEA-Type 2 architecture, which incorporates just one convolutional layer in the encoder section are demonstrated in three test cases.}
    \label{1_layer_autoencoder}
\end{figure}

In an alternative version of MEA-Type 1, only one convolutional layer is used in the encoder segment to increase the number of channels from 1 to 128. After training this MEA architecture (MEA-Type 2), its performance was evaluated using the same test cases that were used for the MEA-Type 1 model with twelve convolutional layers. The results, shown in Fig.~\ref{1_layer_autoencoder}, show mixed results. In some cases, MEA-Type 2 shows acceptable prediction accuracy, but in other cases its performance deteriorates significantly, especially in direct comparison with MEA-Type 1. This discrepancy in accuracy is most evident when comparing error maps, where the model with twelve convolutional layers in the encoder consistently performs better than the single-layer alternative. 

 \begin{figure}[t]
    \raisebox{1ex}{}\hspace{0.14\textwidth}%
    \raisebox{1ex}{Prediction}\hspace{0.06\textwidth}%
    \raisebox{1ex}{FE Results}\hspace{0.03\textwidth}%
    \raisebox{1ex}{Absolute Error}
    \centering    
    \begin{subfigure}{0.65\textwidth}
        \includegraphics[width=\textwidth]{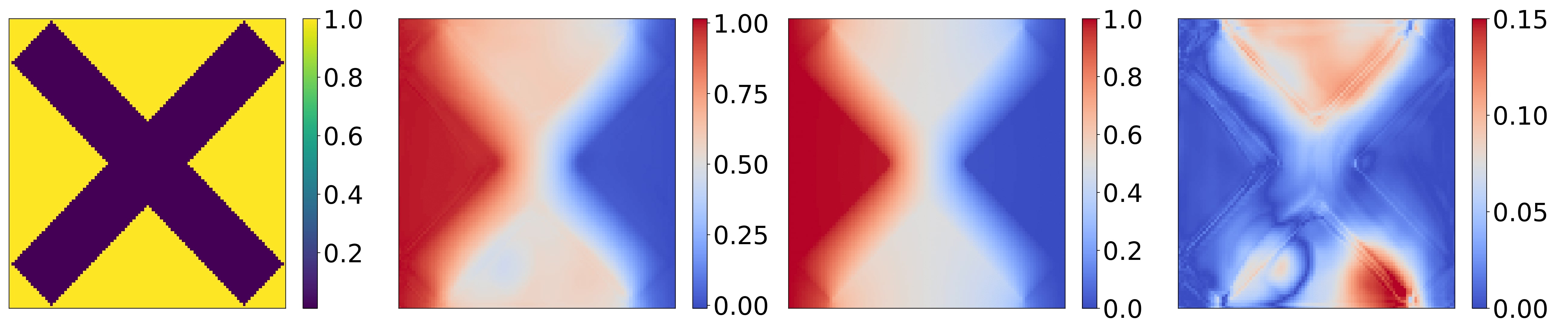}
        \end{subfigure}  
    \begin{subfigure}{0.65\textwidth}
        \includegraphics[width=\textwidth]{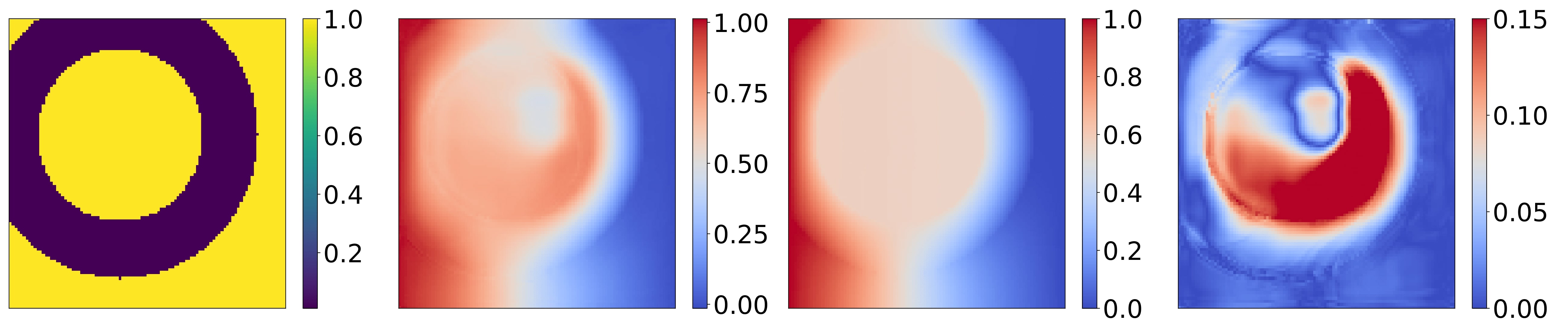}
    \end{subfigure}
    \begin{subfigure}{0.65\textwidth}
        \includegraphics[width=\textwidth]{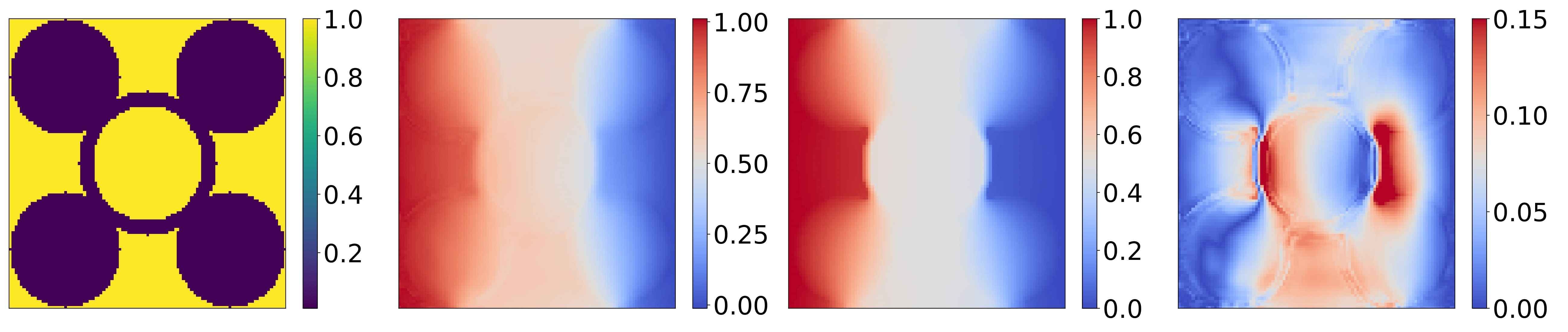}
    \end{subfigure}    
    \caption{Prediction results of the thermal diffusion using standard U-Net architecture section are demonstrated on three test cases. }
    \label{UNet_classic}
\end{figure}

The observed superiority of the MEA network with multiple convolutional layers in the encoder segment over MEA architecture with a single encoder layer can be attributed to an understanding of how deep learning models process and learn from data. Deep convolutional networks learn hierarchically structured features. Initial convolutional layers capture basic aspects like edges and textures, while deeper layers synthesize these into more complex patterns. A single convolutional layer, regardless of its channel depth, lacks this hierarchical structure, potentially limiting its capacity to model complex dependencies and patterns within the data. Besides, the addition of layers increases a network's capacity, allowing it to learn a richer set of features. More convolutional operations in the encoder segment enable the network to capture a broader and more intricate understanding of the data. Hence, this architectural depth enhances the network's robustness to variations and noise in input data, especially critical for accurately predicting OOD test cases. 

The predictive performance of the standard U-Net, as implemented in this study, is showcased in Fig.~\ref{UNet_classic}. This model processes high-resolution microstructures as inputs to produce corresponding high-fidelity solutions as outputs. Examination of the error maps reveals that the standard U-Net outperforms the MEA architecture in certain test cases (e.g., row 2 in Fig.~\ref{9_layer_autoencoder-appandix} and Fig.~\ref{UNet_classic-appandix}), suggesting relatively better prediction accuracy under specific conditions. However, in a wider range of OOD test scenarios, the standard U-Net provides unsatisfactory results in some test cases compared to the MEA models (see Fig.\ref{9_layer_autoencoder} and Fig.\ref{UNet_classic}). These cases show that the standard U-Net does not have the same degree of generalizability as the MEA models with the same amount of training data.
\begin{figure}[t]
\center
    \includegraphics[width=\textwidth]{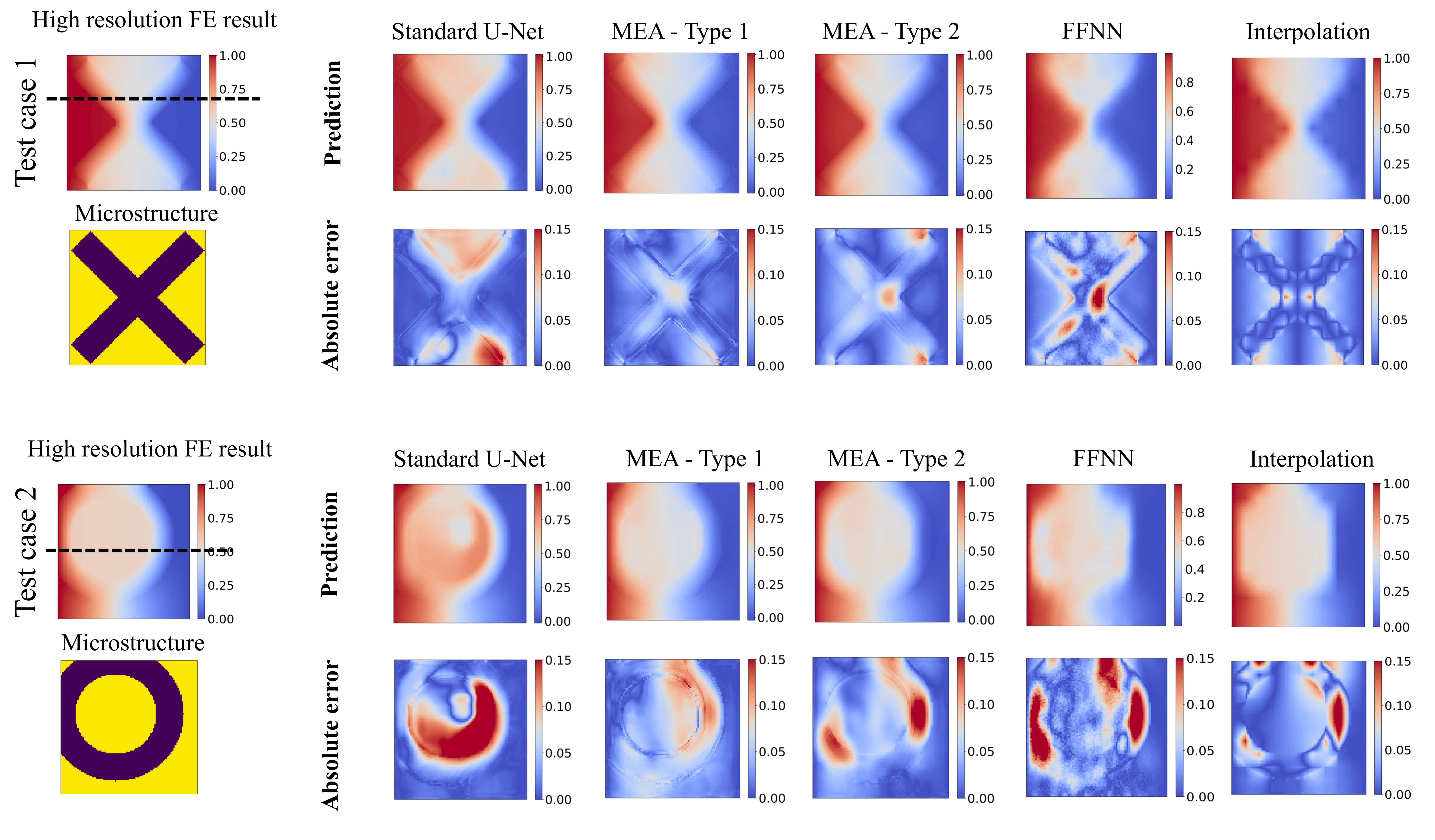}
    \caption{Comparison of the prediction results of the architectures on a single test case. MEA-Type 1 represents an MEA architecture with twelve convolutional layers in the encoder section, while MEA-Type 2 features a model with just one convolutional layer in the encoder segment.  }
    \label{comparison_results}
\end{figure}
\begin{figure}[t]
\center
    \includegraphics[width=\textwidth]{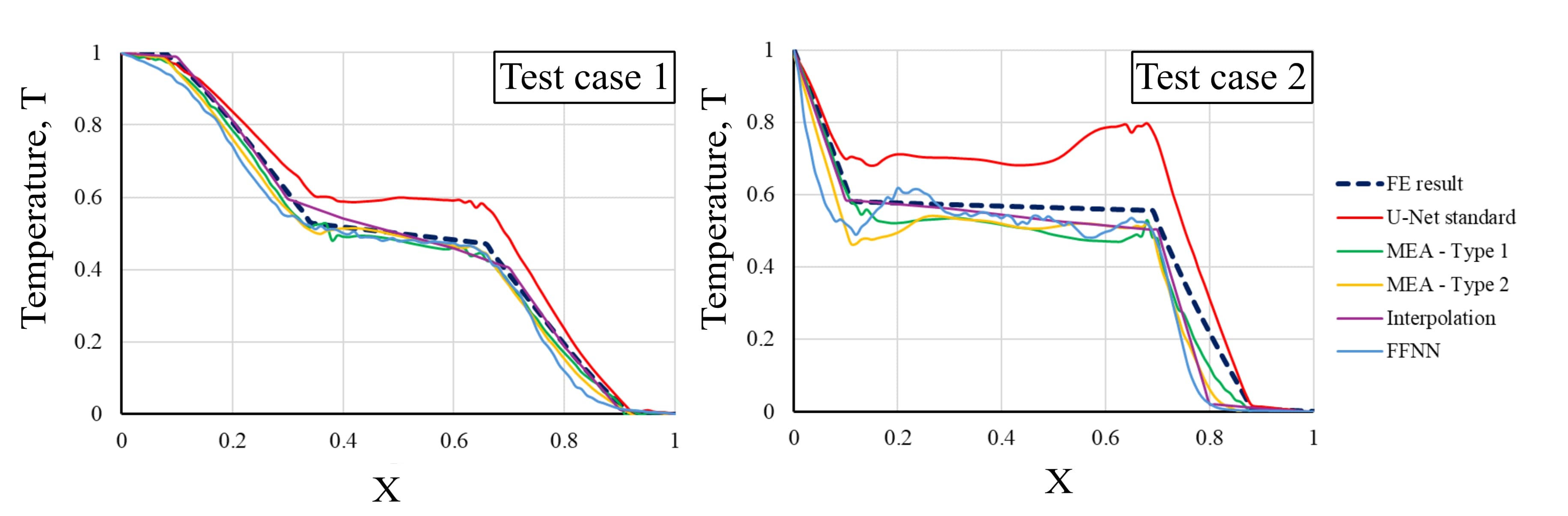}
    \caption{Two-dimensional temperature plot on the cross-section of test cases shown in Fig.~\ref{comparison_results}. The position of the cross-section is indicated by a black dashed line in Fig.~\ref{comparison_results}. The MEA-Type 1 architecture demonstrates accurate predictions when compared to the FE simulation results.}
    \label{cross_section}
\end{figure}

Figure~\ref{comparison_results} presents a comprehensive comparison of the predictions made by various architectures developed in this study, across two test cases. It details the spatial distribution and magnitude of errors for each architecture, highlighting the superior performance of the MEA architectures, particularly MEA-Type1. This advantage is evident in the spatial 2D plot of absolute error values, further corroborated by the statistical analysis in Table~\ref{error_comparison}. This table analyzes the prediction performance across all depicted test cases (see Figure~\ref{test_cases}), starting with the calculation of the mean error for each model on the test cases, followed by computing the average of these mean errors for each model across the six test cases.

Fig.~\ref{cross_section} shows the predicted temperature values that are plotted along the black lines represented in Fig.~\ref{comparison_results}. Besides, the finite element results for each test case are also indicated in Fig.~\ref{cross_section} with dashed lines, showing that less sophisticated architectures like the interpolation model yield results close to the FE results. However, as seen in Fig.~\ref{sharp_interface}, the interpolation model fails to accurately capture the sharp interfaces crucial in fields such as fluid mechanics simulations, rendering it a less suitable choice.

Conversely, the standard U-network model, although capable of identifying sharp interfaces, did not provide promising results in \emph{Test case 2}, as shown in Figure~\ref{cross_section}. However, according to Table~\ref{error_comparison}, the overall performance of the model is comparable to that of MEA-Type 2, although it requires a significantly longer evaluation time. 

Furthermore, Table~\ref{architecture_comparison} presents the total number of parameters for each architecture as well as the evaluation time for a single test case. The test case is uniform across all models to ensure comparability. As observed in Table~\ref{architecture_comparison}, the MEA architectures contain fewer parameters compared to the standard U-Net model, which significantly reduces evaluation time. 
\begin{table}[h]
\centering
\caption{Mean error of the studied architectures on test cases shown in Fig.~\ref{test_cases}. The mean error is calculated by summing the error values for each test case and then dividing by the total number of elements of the corresponding test case.}
\label{error_comparison}
\begin{tabular*}{\textwidth}{@{\extracolsep{\fill}}lccccc}
\toprule
Test cases& Standard U-Net & MEA - Type 1 & MEA - Type 2 & FFNN & Interpolation\\
\midrule
Test 1 & 0.0377  & 0.0226  & 0.0289  & 0.0375 & 0.023 \\

Test 2 & 0.011  & 0.0193  & 0.0203 & 0.0239& 0.023 \\

Test 3 & 0.0056 & 0.0279  & 0.0289  & 0.0313& 0.0178 \\

Test 4 & 0.0528  & 0.0439  &0.0431  & 0.04638& 0.0316 \\

Test 5 & 0.0301  & 0.0220  & 0.0262  & 0.03& 0.0175 \\

Test 6 & 0.0535  & 0.0302  & 0.0356  & 0.0449 & 0.0465 \\
\midrule
\textbf{Average} & \textbf{0.0318} & \textbf{0.0277} & \textbf{0.0305}  & \textbf{0.0357} & \textbf{0.0267}\\
\bottomrule
\end{tabular*}
\end{table}

\begin{table}[h]
\centering
\caption{Comparison of different neural network architectures for obtaining the high-fidelity solution map of a single test case. \textbf{The FE computational cost on the high resolution parametric space of the same test case is equal to 3.619 seconds.}}
\label{architecture_comparison}
\begin{tabular*}{\textwidth}{@{\extracolsep{\fill}}lccccc}
\toprule
 & Standard U-Net & MEA - Type 1 & MEA - Type 2 & FFNN & Interpolation\\
\midrule
Number of parameters & 1,509,145 & 1,226,918 & 1,082,102 & 56,143,201 & - \\
Evaluation time (s) & 3.298 & 0.013 & 0.006 & 0.0419& 0.000995 \\
\bottomrule
\end{tabular*}
\end{table}

\subsection{Training data impact: Standard U-Net vs. MEA approach}
\begin{figure}[t]
\center
    \includegraphics[width=0.8\textwidth]{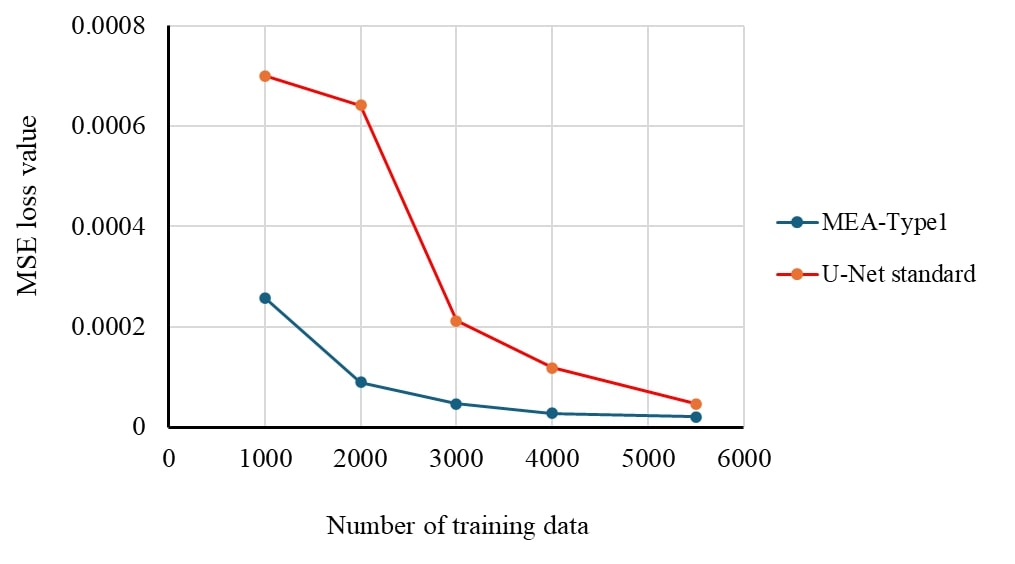}
    \caption{Effect of the amount of training data on the validation loss of standard U-Net and MEA architecture.}
    \label{training_data_graph}
\end{figure}

\begin{figure}[t ]
    \raisebox{1ex}{}\hspace{0.14\textwidth}%
    \raisebox{1ex}{Prediction}\hspace{0.06\textwidth}%
    \raisebox{1ex}{FE Results}\hspace{0.03\textwidth}%
    \raisebox{1ex}{Absolute Error}
    \centering    
    \includegraphics[width=0.7\textwidth]{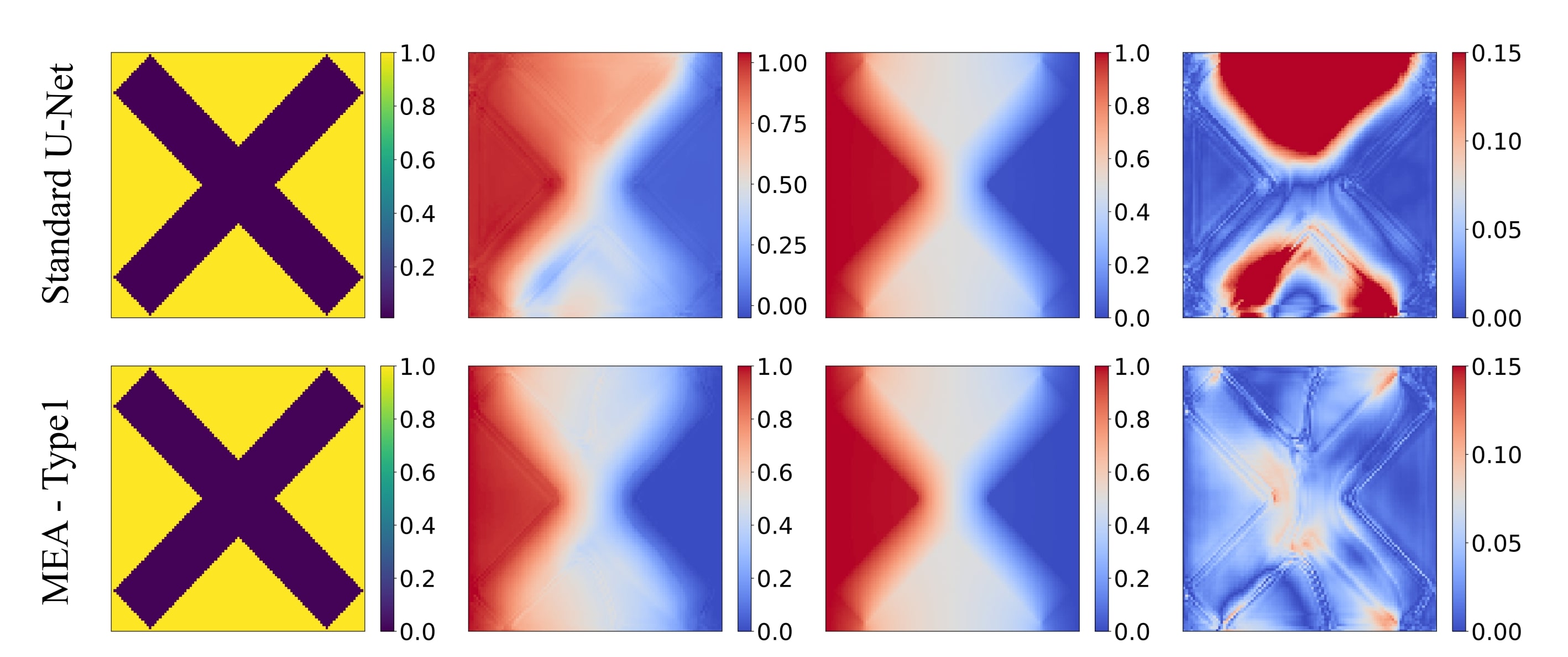}
\caption{Prediction results of the standard U-Net and the MEA-Type 1 model, each trained with only 1000 data points on a single test case. As the absolute error maps show, the MEA model outperforms the standard U-Net, especially when the number of available data sets is limited. A more comprehensive comparison of the models is shown in Fig.~\ref{training_data_effect} across three OOD test cases.}
    \label{training_data_effect1}
\end{figure}

As shown in Fig. ~\ref{training_data_graph}, we analyze the impact of the training data set on two best models in this study, the standard U-Net and the MEA approach. The analysis shows that the MEA architecture clearly outperforms the standard U-Net, especially with limited training data. However, as the amount of training data increases, the performance of the standard U-net converges relatively quickly to that of the MEA model. Therefore, it can be claimed that the MEA model is computationally advantageous when the data available for training is limited. Moreover, the integration possibility of MEA model with other methods such as FOL and FEM to obtain the low-fidelity solution facilitates the development of a multi-fidelity approach, an advantage that the standard U-Net is missing. Figure~\ref{training_data_effect1} shows the prediction results of both models trained with only 1,000 data points, highlighting the superior performance of the MEA model compared to the standard U-Net under the conditions of a small dataset. A more comprehensive comparison of the models is shown in Figure~\ref{training_data_effect} in Appendix A. It can be argued that the combination of a smaller number of parameters, increased computational speed and improved generalizability due to the concatenations features highlights the advantages of the MEA over the conventional U-Net architecture where the amount of data available for training is limited. Considering the desired accuracy level and the available resources, the MEA-Type 2 model can also be preferred due to its significantly shorter evaluation time compared to MEA type 1 (see Table~\ref{architecture_comparison}). However, it should be noted that this efficiency is accompanied by a slight increase in errors.
 
\section{Conclusion}\label{sec5}
In this paper, we introduce the Microstructure-Embedded Autoencoder (MEA), a novel deep learning method developed to upscale low-fidelity solutions into high-fidelity ones. The MEA architecture not only effectively upscales low-resolution solutions to high-resolution ones while preserving critical details, especially discontinuities such as sharp interfaces in the solution field, but also achieves this with significantly lower training data requirements. This reduction in required training data is a major advantage and makes MEA model an efficient and practical solution in this area.

We argue that this method can be effectively integrated into various numerical solvers, including recent developments such as physics-informed neural operators. This capability allows users to perform computations and train neural operators on a reduced space with fewer inputs. Interestingly, we achieved higher accuracy with this approach compared to some existing methods, such as interpolation and the standard U-Net architecture, despite using low amount of training data. The main reason for this improvement is the integration of physical information from the microstructure topology and the inclusion of a novel physically informed operator learning algorithm, both of which are not present in the classical U-Net architecture.

In this work, the proposed method is applied to the thermal diffusion problem within a heterogeneous material and involves three main steps. First, it condenses a high-resolution parametric space representing the spatial distribution of heat conductivity coefficients for two-phase composite materials into low-resolution grids. Second, it solves the boundary value problem on the coarsest grid using a pre-trained physics-informed finite operator model. Finally, it enhances the low-fidelity solution through the MEA architecture to achieve a high-fidelity output. Our findings demonstrate that MEA significantly reduces computational time, costing 280 times less than the FEM method for similar tasks. This reduction in computational cost represents its efficiency and suitability for problems in computational mechanics where both speed and accuracy are crucial. For a summary and comparison of the investigated methods in this paper see also Table \ref{comparison}. 

In future developments, it is worthwhile to extend the methodology to a 3D setting where the advantages and acceleration of the method become even more evident. Additionally, exploring the application of the method to other physical problems, such as mechanical deformation or transient response of structures over time, would be beneficial. Coupling the method with other numerical schemes (e.g. FEM, FFT, and FDM), including alternative neural operators (e.g. DeepOnet and FNO), is also a straightforward and valuable avenue for further investigations.

\begin{table}[h]
    \captionsetup{} 
    \footnotesize 
    \setlength{\tabcolsep}{4pt} 
    \centering
    \caption{Comparison of different methods to obtain a high-fidelity solution map from a given high-fidelity microstructure map. MEA-Type 1 represents an MEA architecture with twelve convolutional layers in the encoder section, while MEA-Type 2 features a model with just one convolutional layer in the encoder segment.}\label{comparison}
    \begin{tabular}{@{}*{6}{P{2.5cm}}@{}} 
        \toprule
        Method & Interpolation approach & Feed Forward Neural Network & MEA, Type 1 ~~~(this work) & MEA, Type 2 ~~~(this work) & Classical U-Net \\
        \hline\hline 
        Computational / training cost & Extremely cheap & Moderate & Cheap & Cheap & Rather high \\
        \midrule
        Implementation complexity & Very easy & Easy & Moderate & Moderate & Moderate \\
        \midrule
        Possible disadvantages & \parbox[t]{2.5cm}{Poor accuracy for sharp solution} & \parbox[t]{2.5cm}{Poor accuracy for sharp solution} & \parbox[t]{2.5cm}{Network complexity} & \parbox[t]{2.5cm}{Network complexity} & \parbox[t]{2.5cm}{Network complexity / training cost} \\
        \midrule
        Main advantage & \parbox[t]{2.5cm}{Easy implementation} & \parbox[t]{2.5cm}{Easy network architecture} & \parbox[t]{2.5cm}{High accuracy even for sharp solution using moderate amount of data} & \parbox[t]{2.5cm}{High accuracy even for sharp solution using moderate amount of data} & \parbox[t]{2.5cm}{Accuracy (upon feeding enough data)} \\
        \bottomrule
    \end{tabular}
\end{table}

\backmatter

\section*{Acknowledgments}
The authors are grateful to the Zentrum für Digitalisierungs- und Technologieforschung der Bundeswehr (dtec.bw)
for their financial support. The author Shahed Rezaei would like to thank the Deutsche Forschungsgemeinschaft
(DFG) for the funding support provided to develop the present work in the project Cluster
of Excellence “Internet of Production” (project: 390621612).

\section*{Declarations}
The authors declare no conflict of interest

\section*{Data availability}
All the source codes and datasets to reproduce results in this study will be openly available on GitHub at \url{https://github.com/RasoulNajafi/Microstructure-Embedded-Autoencoder-MEA-/tree/main}.

\begin{appendices}

\newpage
\section{Additional Results}\label{secA1}
In this section, we present the prediction results of the implemented MEA architectures and the standard U-Net across all six test cases. In addition, the predictive capabilities of the investigated models in capturing the gradient terms in heat flux problems are analyzed, and the prediction results of the heat flux are compared with the FE simulation results. 
\begin{figure}[H]
    \raisebox{1ex}{}\hspace{0.14\textwidth}%
    \raisebox{1ex}{Prediction}\hspace{0.06\textwidth}%
    \raisebox{1ex}{FE Results}\hspace{0.03\textwidth}%
    \raisebox{1ex}{Absolute Error}
    \centering    
    \begin{subfigure}{0.65\textwidth}
        \includegraphics[width=\textwidth]{figs/UNet_V3_type1-1.jpg}
        \end{subfigure}
    
    \begin{subfigure}{0.65\textwidth}
        \includegraphics[width=\textwidth]{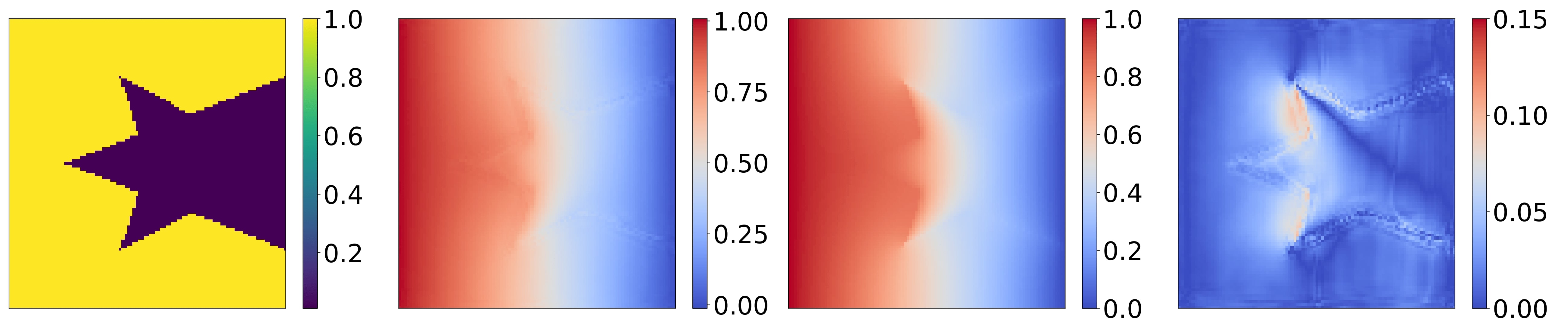}
    \end{subfigure}
    
    \begin{subfigure}{0.65\textwidth}
        \includegraphics[width=\textwidth]{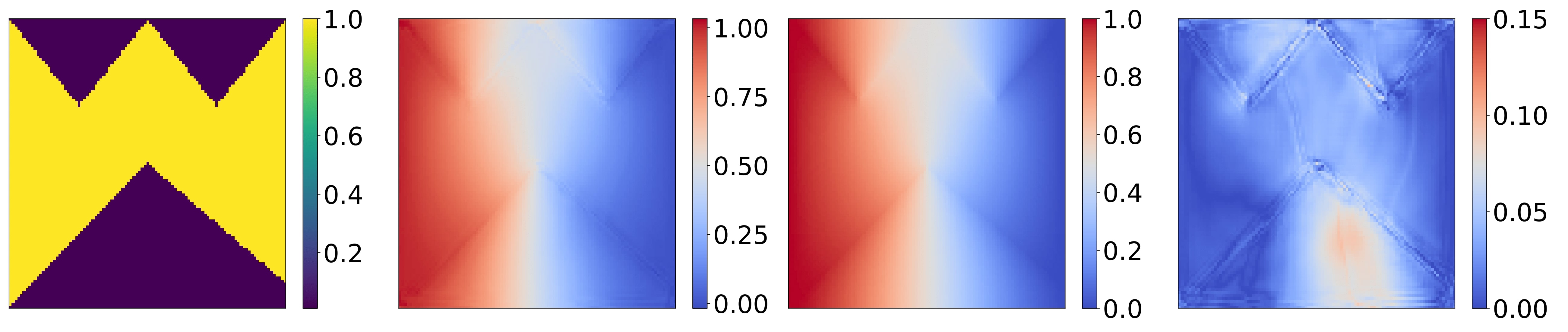}
    \end{subfigure}
    
    \begin{subfigure}{0.65\textwidth}
        \includegraphics[width=\textwidth]{figs/UNet_V3_type1-4.jpg}
    \end{subfigure}
    
    \begin{subfigure}{0.65\textwidth}
        \includegraphics[width=\textwidth]{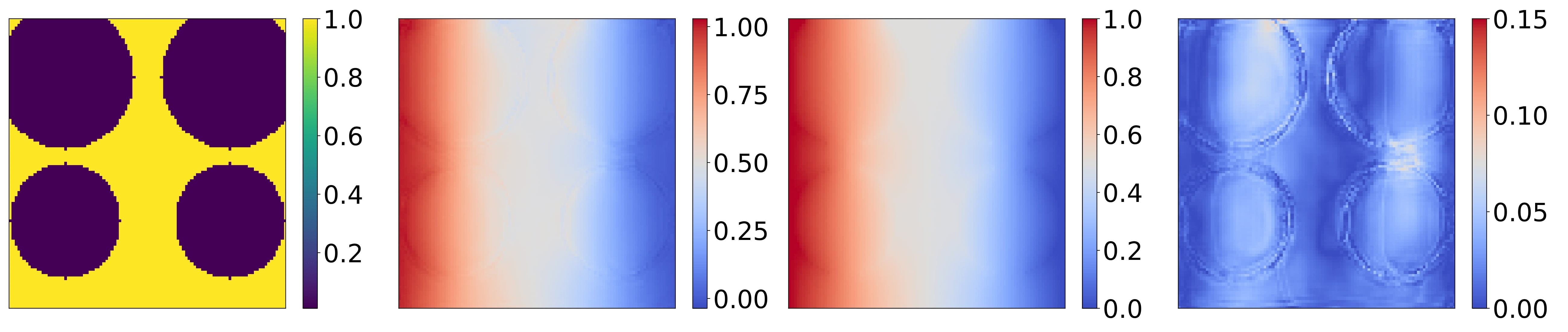}
    \end{subfigure}
    
    \begin{subfigure}{0.65\textwidth}
        \includegraphics[width=\textwidth]{figs/UNet_V3_type1-6.jpg}
    \end{subfigure}    
    \caption{Prediction results of the MEA-Type1 architecture, which incorporates twelve convolutional layers in the encoder section to convert low-fidelity solutions of the steady state heat transfer problem into high-fidelity ones, are demonstrated on six test cases.}
    \label{9_layer_autoencoder-appandix}
\end{figure}

\begin{figure}[H]
    \raisebox{1ex}{}\hspace{0.14\textwidth}%
    \raisebox{1ex}{Prediction}\hspace{0.06\textwidth}%
    \raisebox{1ex}{FE Results}\hspace{0.03\textwidth}%
    \raisebox{1ex}{Absolute Error}
    \centering   
    \begin{subfigure}{0.65\textwidth}
        \includegraphics[width=\textwidth]{figs/UNet_V3_type3-1.jpg}
        \end{subfigure}
    
    \begin{subfigure}{0.65\textwidth}
        \includegraphics[width=\textwidth]{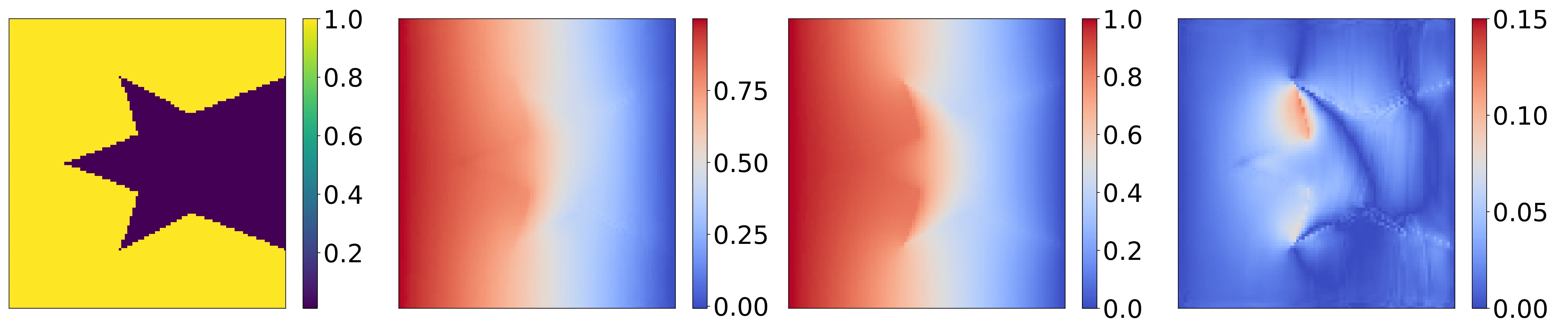}
    \end{subfigure}
    
    \begin{subfigure}{0.65\textwidth}
        \includegraphics[width=\textwidth]{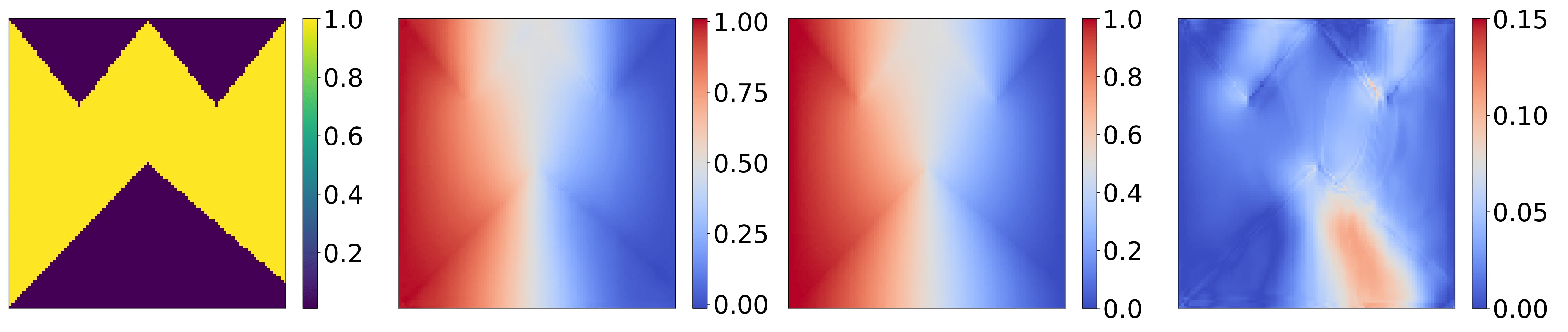}
    \end{subfigure}
    
    \begin{subfigure}{0.65\textwidth}
        \includegraphics[width=\textwidth]{figs/UNet_V3_type3-4.jpg}
    \end{subfigure}
    
    \begin{subfigure}{0.65\textwidth}
        \includegraphics[width=\textwidth]{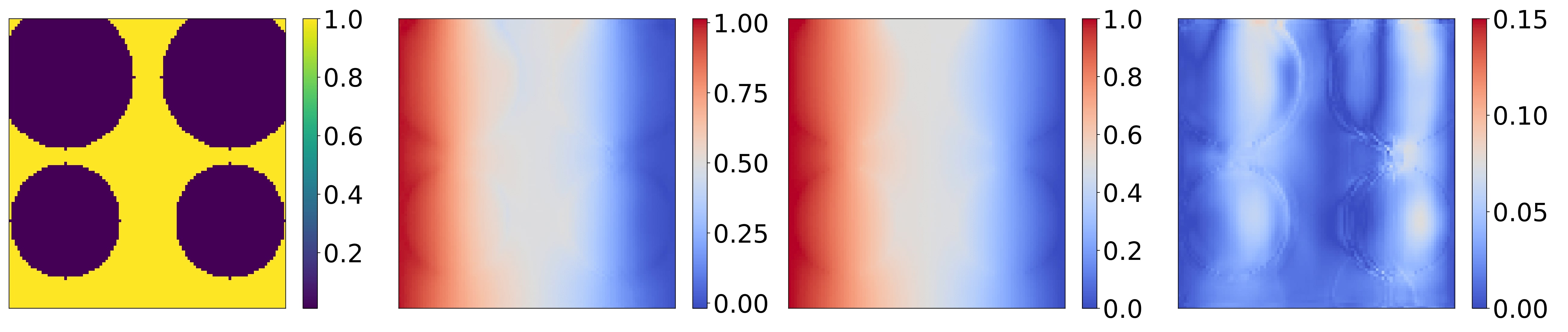}
    \end{subfigure}
    
    \begin{subfigure}{0.65\textwidth}
        \includegraphics[width=\textwidth]{figs/UNet_V3_type3-6.jpg}
    \end{subfigure}
    
    \caption{Prediction results of enhanced MEA-Type2 architecture, which incorporates just one convolutional layer in the encoder section are demonstrated on six test cases.}
    \label{1_layer_autoencoder-appandix}
\end{figure}

 \begin{figure}[H]
    \raisebox{1ex}{}\hspace{0.14\textwidth}%
    \raisebox{1ex}{Prediction}\hspace{0.06\textwidth}%
    \raisebox{1ex}{FE Results}\hspace{0.03\textwidth}%
    \raisebox{1ex}{Absolute Error}
    \centering    
    \begin{subfigure}{0.65\textwidth}
        \includegraphics[width=\textwidth]{figs/UNet_SH_classic-1.jpg}
        \end{subfigure}
    
    \begin{subfigure}{0.65\textwidth}
        \includegraphics[width=\textwidth]{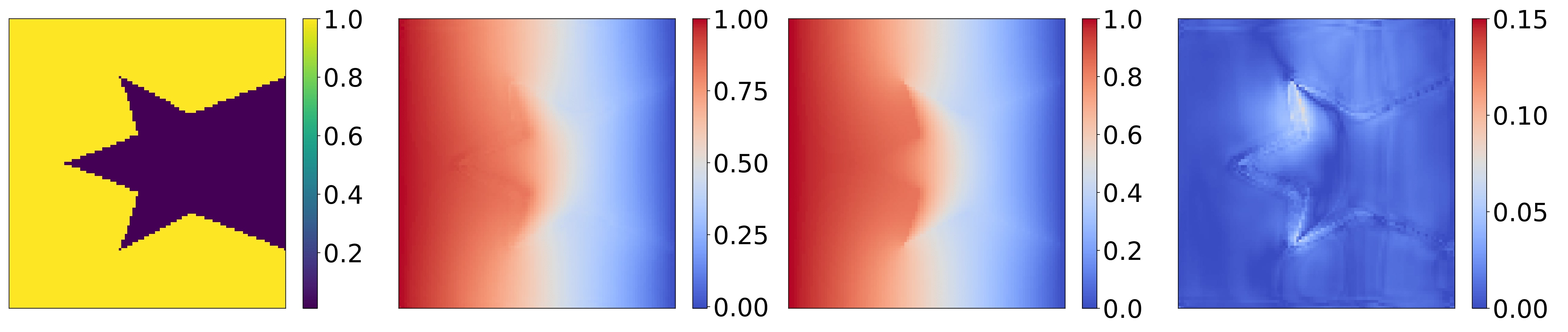}
    \end{subfigure}
    
    \begin{subfigure}{0.65\textwidth}
        \includegraphics[width=\textwidth]{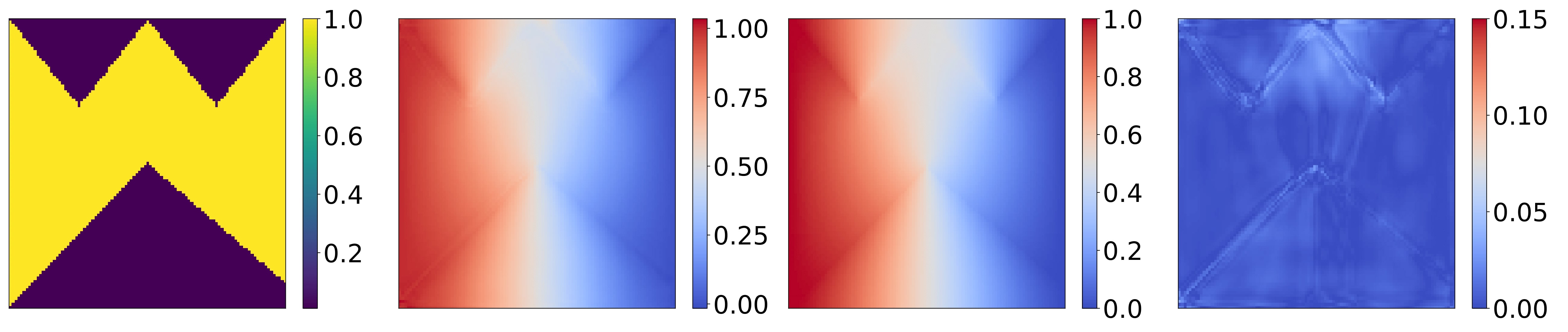}
    \end{subfigure}
    
    \begin{subfigure}{0.65\textwidth}
        \includegraphics[width=\textwidth]{figs/UNet_SH_classic-4.jpg}
    \end{subfigure}
    
    \begin{subfigure}{0.65\textwidth}
        \includegraphics[width=\textwidth]{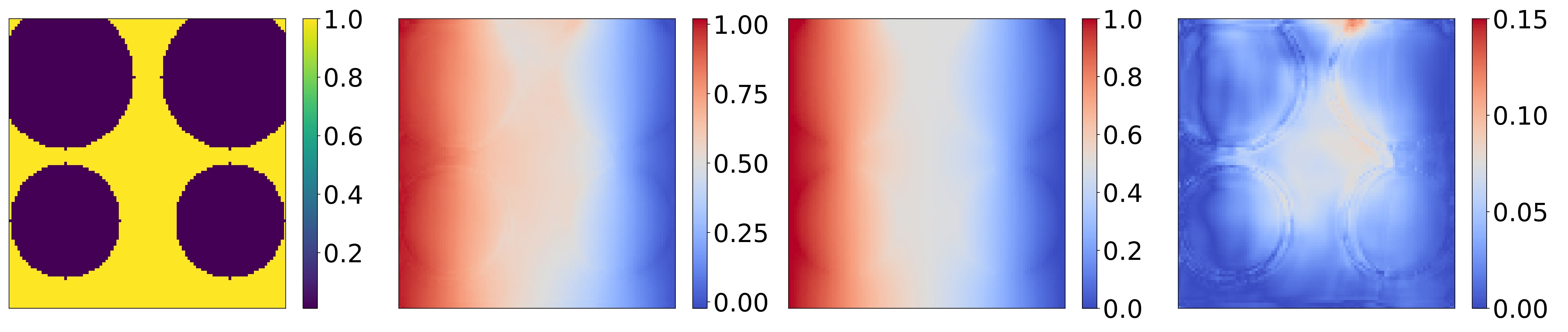}
    \end{subfigure}
    
    \begin{subfigure}{0.65\textwidth}
        \includegraphics[width=\textwidth]{figs/UNet_SH_classic-6.jpg}
    \end{subfigure}    
    \caption{Prediction results of the standard U-Net architecture section are demonstrated on six test cases. }
    \label{UNet_classic-appandix}
\end{figure}

\begin{figure}[H]
    \raisebox{1ex}{}\hspace{0.14\textwidth}%
    \raisebox{1ex}{Prediction}\hspace{0.06\textwidth}%
    \raisebox{1ex}{FE Results}\hspace{0.03\textwidth}%
    \raisebox{1ex}{Absolute Error}
    \centering    
    \begin{subfigure}{0.65\textwidth}
        \includegraphics[width=\textwidth]{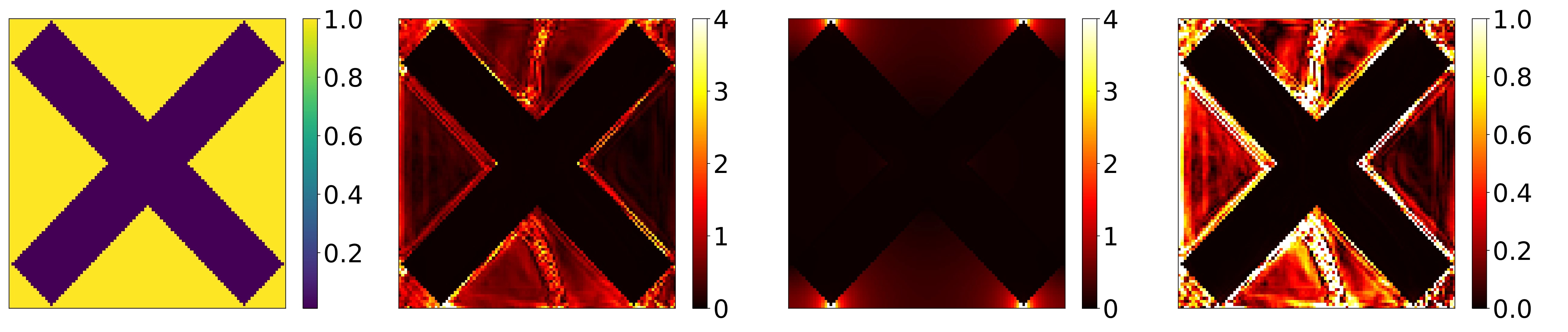}
        \end{subfigure}    
    \begin{subfigure}{0.65\textwidth}
        \includegraphics[width=\textwidth]{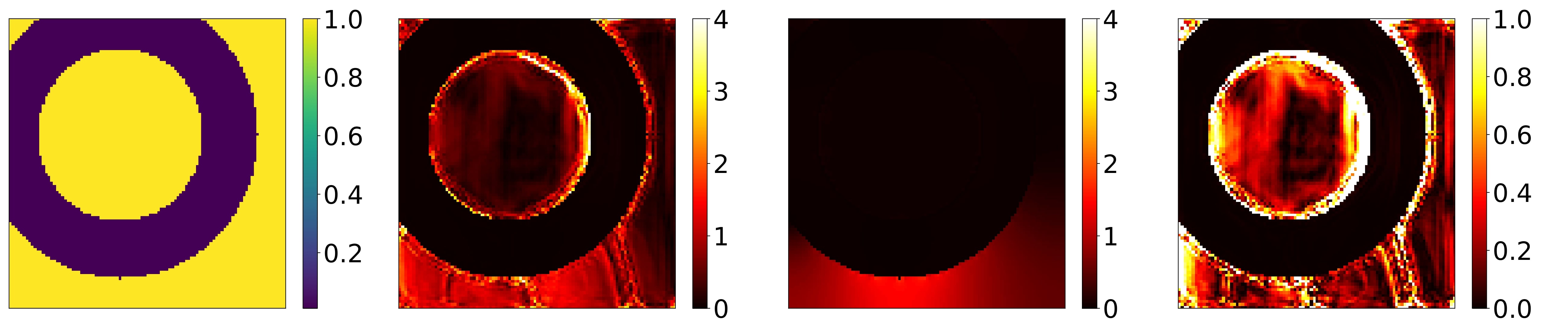}
    \end{subfigure}    
    \begin{subfigure}{0.65\textwidth}
        \includegraphics[width=\textwidth]{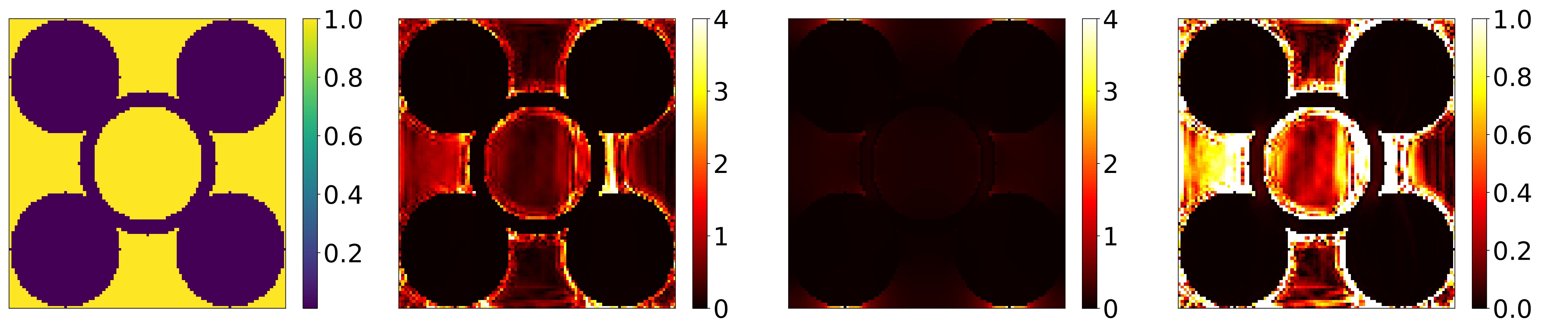}
    \end{subfigure}    
    \caption{Prediction results of heat flux using the MEA-Type1 model, featuring a twelve convolutional operation in the encoder section.}
    \label{9_layer_Flux-appandix}
\end{figure}

\begin{figure}[H]
    \raisebox{1ex}{}\hspace{0.14\textwidth}%
    \raisebox{1ex}{Prediction}\hspace{0.06\textwidth}%
    \raisebox{1ex}{FE Results}\hspace{0.03\textwidth}%
    \raisebox{1ex}{Absolute Error}
    \centering    
    \begin{subfigure}{0.65\textwidth}
        \includegraphics[width=\textwidth]{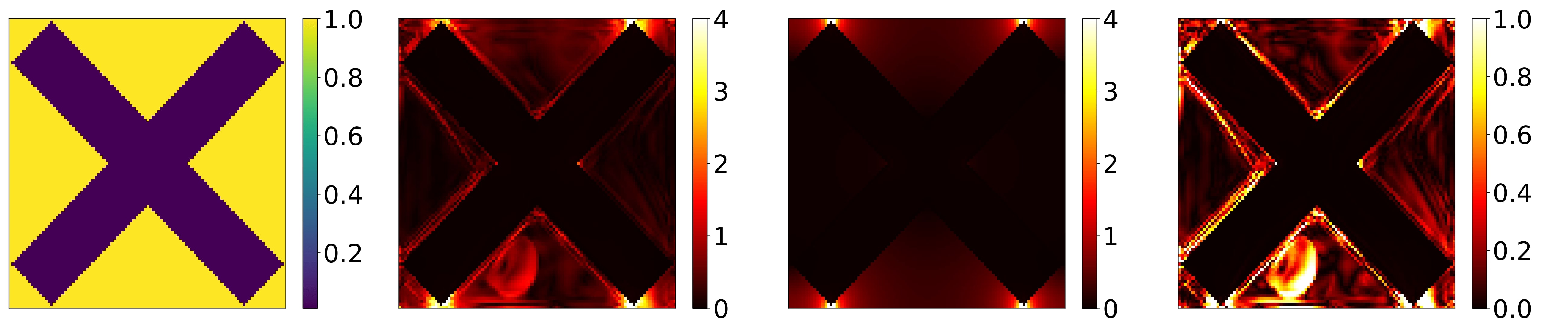}
        \end{subfigure}
    
    \begin{subfigure}{0.65\textwidth}
        \includegraphics[width=\textwidth]{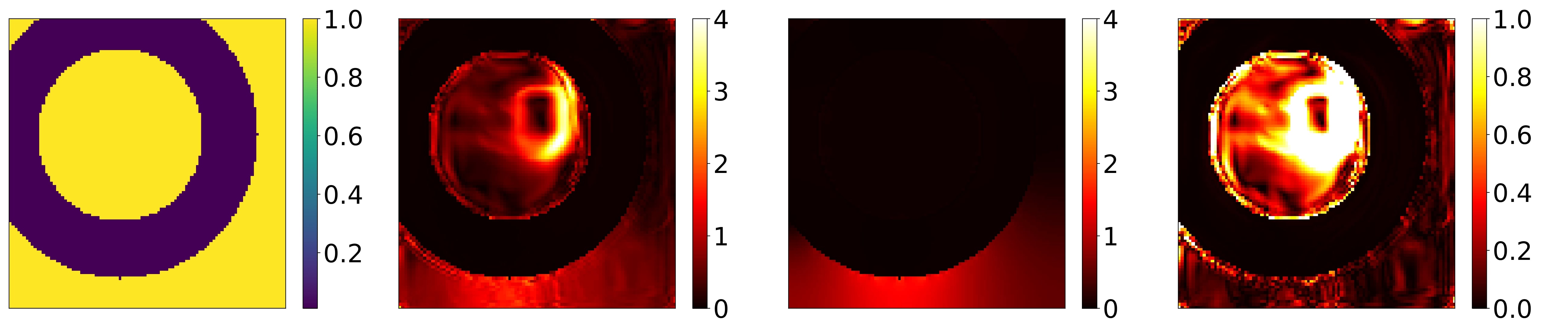}
    \end{subfigure}
    
    \begin{subfigure}{0.65\textwidth}
        \includegraphics[width=\textwidth]{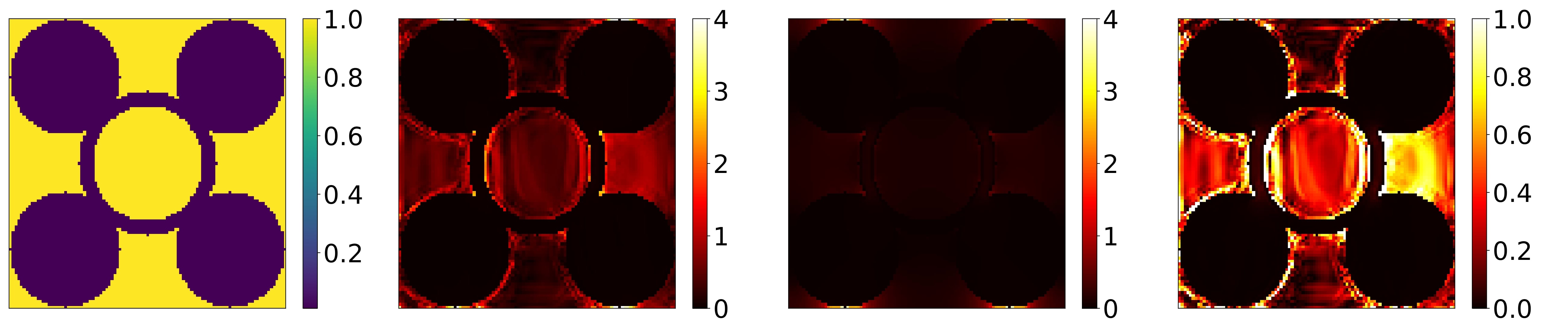}
    \end{subfigure}   
    \caption{Prediction results of heat flux using the standard U-Net architecture across three test scenarios.}
    \label{UNet_classic_Flux-appandix}
\end{figure}

To further evaluate the performance of the MEA architectures and compare them with the standard U-Net, we conducted an analysis focusing on their ability to capture gradient terms after predicting spatial temperature distributions. This analysis involves calculating the gradient of the predicted temperature first and then multiplying it with the spatial heat conductivity coefficients to determine the heat flux. By comparing Fig.~\ref{9_layer_Flux-appandix} which illustrates the results from the MEA-Type1 architectures with the results from the standard U-Net shown in Figure~\ref{UNet_classic_Flux-appandix}, it is clear that the result of both methods deviates from the finite element calculations. However, the initial focus of this study is on predicting the primary variable, which in this case is the temperature field. Future work will need to expand on this to improve predictions for the spatial derivatives of the primary variable, as demonstrated in \cite{REZAEI2022PINN}.

\end{appendices}

\begin{figure}[H]
    \raisebox{1ex}{}\hspace{0.14\textwidth}%
    \raisebox{1ex}{Prediction}\hspace{0.06\textwidth}%
    \raisebox{1ex}{FE Results}\hspace{0.03\textwidth}%
    \raisebox{1ex}{Absolute Error}
    \centering    
    \includegraphics[width=0.65\textwidth]{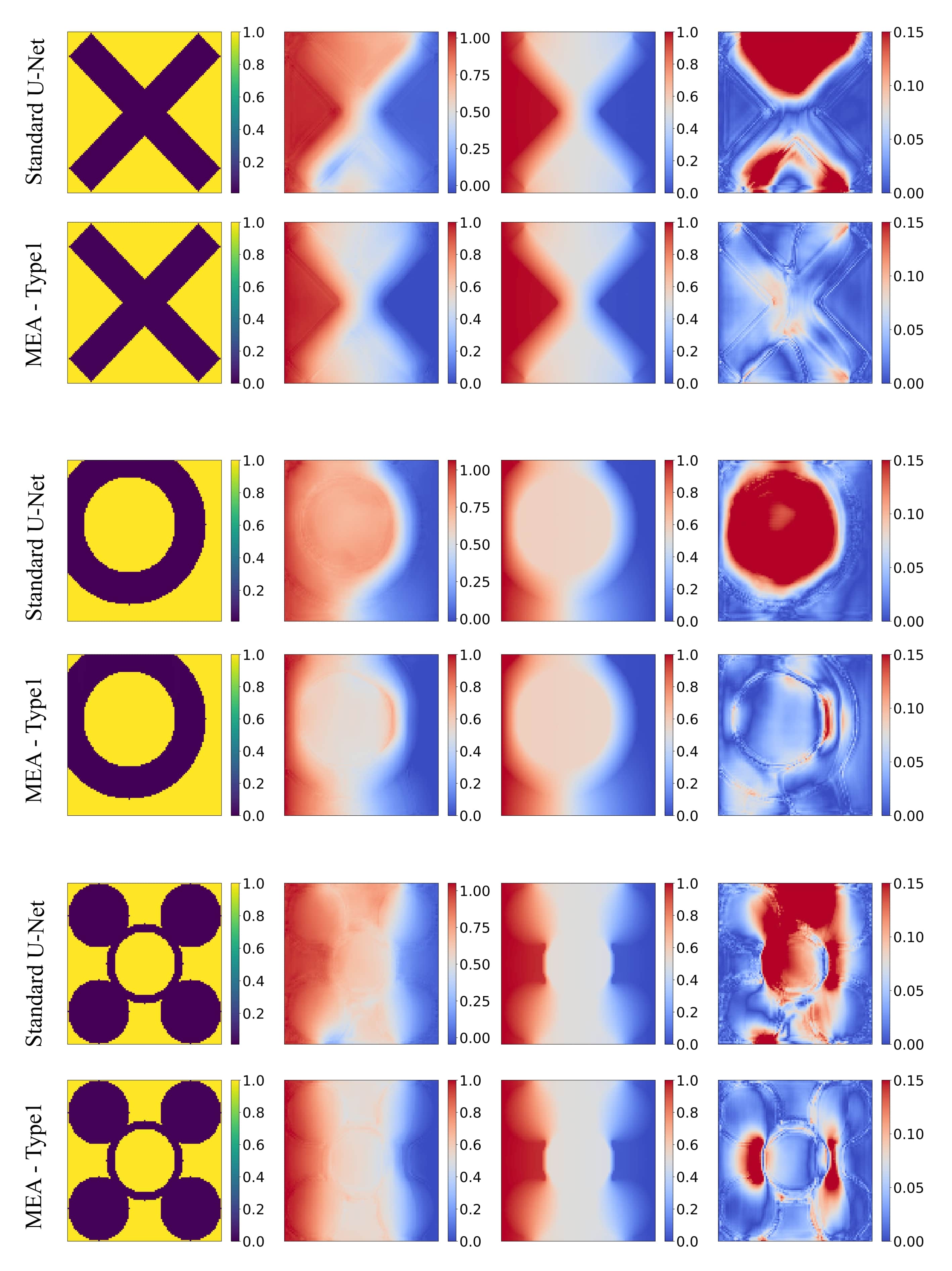}
    \caption{Prediction results of the standard U-Net and MEA-Type 1 model for three test cases. Both models are trained with 1,000 data points. As can be observed, the MEA architecture shows better performance in terms of accuracy compared to the standard U-Net when the amount of training data points is considerably small. }
    \label{training_data_effect}
\end{figure}

\end{document}